\documentclass[12pt, a4]{article}

\usepackage{graphicx}
\usepackage{amsmath}
\usepackage{amssymb}

\newcommand{\la}{\lambda}
\newcommand{\al}{\alpha}

\newcommand{\sn}{\mathrm{sn}}

\newcommand{\prt}{\partial}

\begin{document}

\title{Evolution of solitary waves and undular bores in  shallow-water
flows  over a gradual slope with bottom friction}

\author{G.A.~El$^1$, \ R.H.J.~Grimshaw$^2$  \\
 Department of Mathematical Sciences, Loughborough University,\\
Loughborough LE11 3TU, UK \\
$^1$ e-mail: G.El@lboro.ac.uk \ \ \ $^2$ e-mail:
R.H.J.Grimshaw@lboro.ac.uk
 \\
 \\
A.M.~Kamchatnov \\
Institute of Spectroscopy, Russian Academy of Sciences,\\
Troitsk, Moscow Region,  142190 Russia \\
e-mail: kamch@isan.troitsk.ru}
\date{}
\maketitle

\begin{abstract}
This paper considers the propagation of  shallow-water solitary and
nonlinear periodic waves over a gradual slope with bottom friction
in the framework of a variable-coefficient Korteweg-de Vries
equation.
 We use the Whitham averaging method, using a recent
development of this theory for  perturbed integrable equations. This
general approach enables us not only to improve known results on
the adiabatic evolution of isolated solitary waves and periodic
wave trains  in the presence of variable topography and
bottom friction, modeled by the Chezy law,  but also importantly,
to study the effects of these factors on the propagation of undular bores,
which are essentially unsteady in the system under consideration.
In particular, it is shown that the
combined action of variable topography and bottom friction generally
imposes certain global restrictions on the undular bore propagation
so that the evolution of the  leading solitary wave can be substantially
different from that of an isolated solitary wave with the same initial
amplitude. This non-local effect is due to nonlinear wave
interactions within the undular bore and can lead to an additional
solitary wave amplitude growth, which cannot be predicted in the
framework  of the traditional adiabatic approach to the  propagation of
solitary waves in slowly varying media.

\end{abstract}

\section{Introduction}


There have been many studies of the propagation of water waves over a slope,
sometimes also subject to the effects of bottom friction. Many of these works have considered
linear waves, or have been numerical simulations in the framework of various
nonlinear long-wave model equations. Our interest here is in the propagation of
weakly nonlinear long water waves over a slope, simultaneously subject to bottom friction,
a combination apparently first considered by Miles (1983a,b) albeit for the special
case of a single solitary wave, or a periodic wavetrain.
An appropriate model equation for this scenario is the variable-coefficient perturbed
Korteweg-de Vries  (KdV) equation
(see Grimshaw 1981, Johnson 1973a,b),
\begin{equation}\label{1-1}
    A_t+cA_x+\frac{c_x }{2}A+\frac{3c}{2h}AA_x+ \frac{ch^2 }{6} A_{xxx}
    \,=\, -C_D\frac{c}{h^2 }|A|A.
\end{equation}
Here   $A(x,t)$ is the free surface elevation above the undisturbed
depth $h(x)$ and  $c(x)=\sqrt{gh(x)}$ is the linear long wave phase
speed.   The bottom friction term on the right-hand side is
represented by the Chezy law, modelling a turbulent boundary layer.
Here  $C_D$ is a non-dimensional drag coefficient, often assumed to
have a value around   $0.01 $ (Miles 1983a,b). Other forms of
friction could be used (see, for instance Grimshaw et al 2003) but
the Chezy law seems to be the most appropriate for water waves in a
shallow depth. In (\ref{1-1}) the first two terms on the left-hand
side are the dominant terms, and by themselves describe the
propagation of a linear long wave with speed $c$. The remaining
terms on the left-hand side represent, respectively, the effect of
varying depth, weakly nonlinear effects and weak linear dispersion.
The equation is derived using the usual KdV balance in which the
linear dispersion, represented by $\partial^2 /\partial x^2 $, is
balanced by nonlinearity, represented by $A$. Here we have added to
this balance weak inhomogeneity so that $c_x/c $ scales as $h^2
\partial^3 /\partial x^3 $, and weak friction so that $C_D  $ scales
with $h\partial /\partial x $. Within this basic balance of terms,
we can cast  (\ref{1-1})  into the asymptotically equivalent form
\begin{equation}\label{1-2}
    A_{\tau}+\frac{h_{\tau }}{4h}A+\frac{3}{2h}AA_X+\frac{h}{6g}A_{XXX}=
    -C_D\frac{g^{1/2}}{h^{3/2}}|A|A,
\end{equation}
\begin{equation}\label{1-3}
\hbox{where} \quad     \tau=\int^{x}_{0}\frac{dx'}{c(x')},\qquad
X=\tau-t.
\end{equation}
Here  we have $h=h(x(\tau))$, explicitly dependent on the variable
$\tau $ which describes evolution along the path of the wave.

The governing equation (\ref{1-2}) can be cast into several equivalent forms.
That most commonly used is the variable-coefficient KdV equation,
obtained here by putting
\begin{equation}\label{B}
B=(gh)^{1/4}A
\end{equation}
\begin{equation}\label{kdvB}
\hbox{so that} \quad B_{\tau}+\frac{3}{2g^{1/4}h^{5/4}}BB_X+\frac{h}{6g}B_{XXX} \,=\,
-C_D\frac{g^{1/4}}{h^{7/4}}|B|B  \, .
\end{equation}
This form shows that, in the absence of friction term, i.e. when
$C_D\equiv 0$, equation (\ref{1-2}) has two integrals of motion
with the densities proportional to $h^{1/4}A$ and $h^{1/2}A^2$.
These are often referred to as laws for the conservation of ``mass''
and ``momentum''.  However, these densities do not
necessarily correspond to the corresponding physical entities.  Indeed,
to leading order,  the ``momentum'' density is proportional to the
wave action flux, while the ``mass'' density differs slightly from  the actual
mass density. This latter  issue has been explored by Miles (1979),
where it was shown that the difference is smaller than
the error incurred in the derivation of  equation (\ref{B}),
and is due to reflected waves.

Our main concern in this paper is with the behaviour of an undular
bore over a slope in the presence of bottom friction, using the
perturbed KdV equation (\ref{1-2}), where we were originally
motivated by the possibility that the behaviour of a tsunami
approaching the shore might be modeled in this way.  The undular
bore solution to the unperturbed KdV equation  can be constructed
using the well-known Gurevich-Pitaevskii (GP) (1974) approach (see
also Fornberg and Whitham 1978). In this approach, the undular bore
is represented as a modulated nonlinear periodic wave train. The
main feature of this unsteady undular bore is the presence of a
solitary wave (which is the limiting wave form of the periodic
cnoidal wave) at its leading edge. The original initial-value
problem for the KdV equation is then replaced by a certain
boundary-value problem for the associated modulation Whitham
equations. We note, however, that so far, the simplest,
``$(x/t)$''-similarity solutions of the modulation equations have
been used for the  modelling of  undular bores in various contexts
(see Grimshaw and Smyth 1986, Smyth 1987 or Apel 2003 for instance).
These solutions, while effectively describing many features of
undular bores, are degenerate and fail to capture, even
qualitatively, some important effects associated with
non-self-similar modulation dynamics. In particular, in the
classical GP solution for the resolution of an initial jump in the
unperturbed KdV equation, the amplitude of the lead solitary wave in
the undular bore is constant (twice the value of the initial jump).
On the other hand, the modulation solution for the undular bore
evolving from a general monotonically decreasing initial profile
shows that the lead solitary wave amplitude in fact grows with time
(Gurevich, Krylov and Mazur 1989; Gurevich, Krylov and El 1992;
Kamchatnov 2000). As we shall see, the very possibility of such
variations in the modulated solutions of the unperturbed KdV
equation has a very important fluid dynamics implication: {\it in a
general setting, the undular bore lead solitary wave cannot be
treated as an individual KdV solitary wave but rather represents a
part of the global nonlinear wave structure}. In other words, while
at every particular moment of time the lead solitary wave has the
spatial profile of the familiar KdV soliton, generally, the temporal
dependence of its amplitude cannot be obtained in the framework  of
single solitary wave perturbation theory.

In the unperturbed KdV equation, the growth of the lead solitary
wave amplitude is caused by the spatial inhomogeneity of the initial
data. Here, however, the presence of a perturbation due to topography and/or friction
serves as an alternative and/or additional cause  for  variation of the lead solitary
wave amplitude. Thus, in the present case, the variation in the amplitude
 will  have two components  (which generally, of course, cannot be
separated because of the nonlinear nature of the problem);  one is
local, described by the adiabatic perturbation theory for a single
solitary wave, and the other one is nonlocal, which in principle
requires the study of the full modulation solution. Depending on the
relative values of the small parameters associated with the slope,
friction and spatial non-uniformity of the initial modulations,  we
can take into account only one of these components, or a combination
of them.

The structure of the paper is as follows. First, in Section 2, we
reformulate the basic model (\ref{1-1}) as a constant-coefficient
KdV equation perturbed by terms representing topography and
friction. Then we derive in Section 3 the associated perturbed
Whitham modulation equations using  methods recently developed by
Kamchatnov (2004). Next, in Section 4, this Whitham system is
integrated in the solitary-wave limit.  Our purpose here is
primarily to obtain the equation of a multiple characteristic, which
defines
 the leading edge of a shoaling undular
bore in the case when the modulations due to the combined action of
the slope and bottom friction are small compared to the existing
spatial modulations due to non-uniformity of the initial data. As a
by-product of this integration, we reproduce and extend the known
results on the adiabatic variation of a single solitary wave (Miles
1983a,b). Then, in Section 5, we carry out an analogous study of a
cnoidal wave, propagating over a gradual slope and subject to
friction, a case studied previously by Miles (1983 b) but under the
restriction of zero mean flow, which is removed here. Finally, in
Section 6 we study effects of a gradual slope and bottom friction on
the front of an undular bore which represents a modulated cnoidal
wave transforming into a system of weakly interacting solitons near
its leading edge.

\section{Problem formulation}

For the purpose of the present paper it is convenient to recast
(\ref{1-2}) into the standard KdV equation form  with constant
coefficients, modified by certain  perturbation terms. Thus we
introduce the new variables\\
\begin{equation}\label{newvar}
    U=\frac{3g}{2h^2}A,\qquad T=\frac1{6g}\int^{\tau}_{0} h d\tau =
\frac{1}{6g^{3/2} }\int_{0}^x \sqrt{h(x)} dx.
\end{equation}
\begin{equation}\label{U}
\hbox{so that} \quad U_T+6UU_X+U_{XXX}=R=F(T)U-G(T)|U|U,
\end{equation}
\begin{equation}\label{F}
\hbox{where} \quad    F(T)=-\frac{9h_T}{4h},\qquad G(T)=4C_D\frac{g^{1/2}}{h^{1/2}}.
\end{equation}
In this form, the governing equation (\ref{U}) has the structure of
the integrable KdV equation on the left-hand side, while the
separate effects of the varying depth and the bottom friction are
represented by the two terms on the right-hand side. This structure
enables us to use the general theory developed in Kamchatnov (2004)
for perturbed integrable systems.

For much of the subsequent discussion, it is useful to assume
that $h(x)= \hbox{constant}$, $C_D=0$ for $x<0$ in the original
equation (\ref{1-1}), which corresponds to $F(T)=G(T)=0$ for $T<0$
in (\ref{U}).  We shall also assume that  $A = 0$ for $x>0$ at
$t=0$, which corresponds to $U=0$ for $X>0$ on $X=\tau(T)$
(see (\ref{newvar})).  Then we shall propose two types of
initial-value problem for (\ref{1-1}), and correspondingly for (\ref{U}).

(a) Let a solitary wave of a given amplitude $a_0$ initially
propagating over a flat bottom without friction (i.e a soliton
described by an  unperturbed KdV equation),  enter the variable
topography and bottom friction region at $t=0$, $x=0$ (Fig. 1 a).

(b) Let an undular bore of a given intensity propagate over a flat
bottom without friction (the corresponding solution of the unperturbed
KdV equation will be discussed in Section 5). Let the lead solitary
wave of this undular bore have the same amplitude $a_0$ and enter
the variable topography and bottom friction region at
$t=0$, $x=0$ (Fig. 1b).

\begin{figure}[bt]
\begin{center}
\includegraphics[width=8cm,height= 2.9cm, clip]{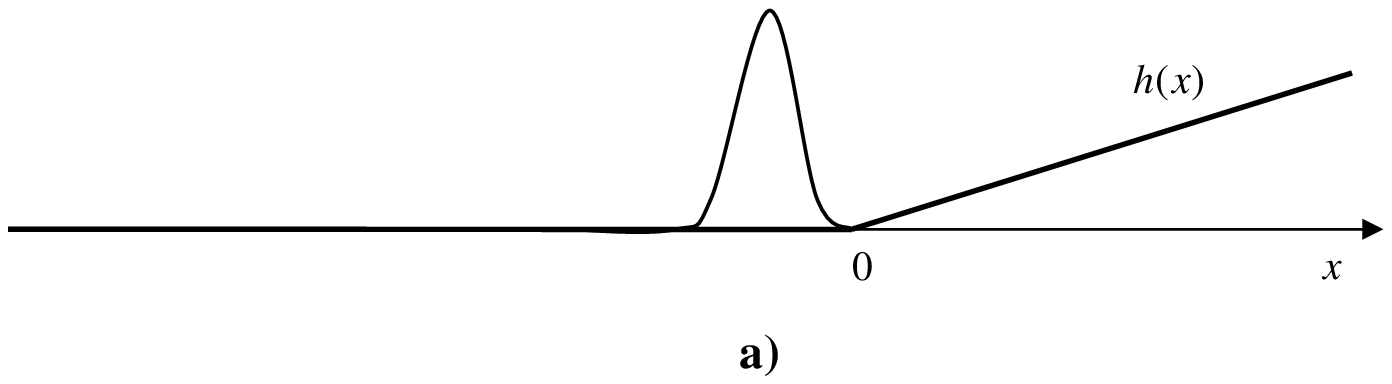} \\ \vspace{1.0cm}
\includegraphics[width=8cm,height= 3.5cm,clip]{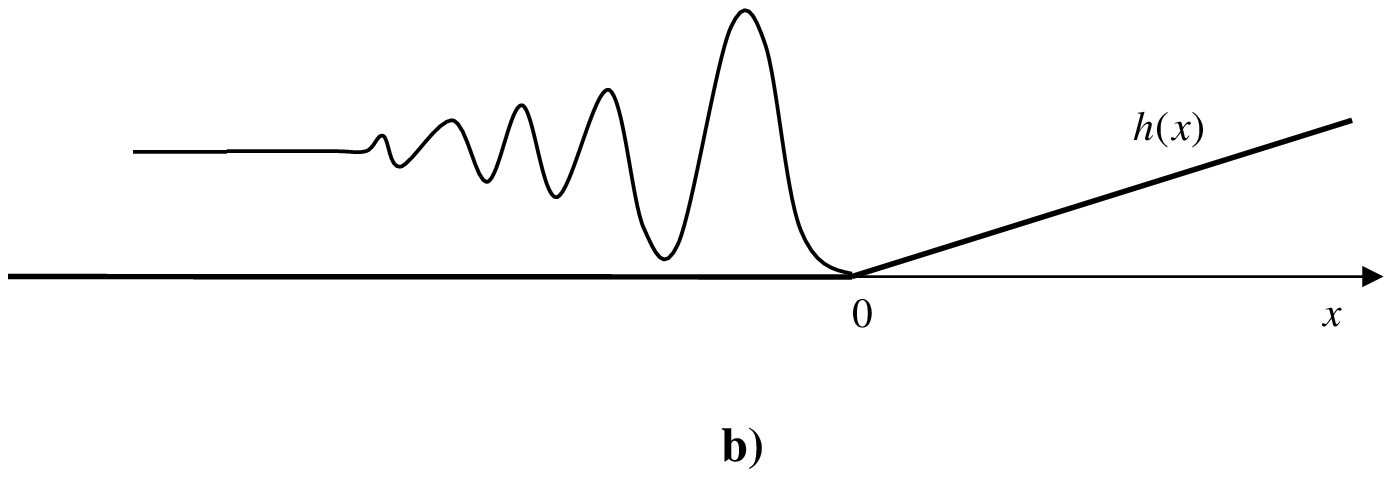}
\end{center}
\caption{Isolated solitary wave (a) and undular bore (b) entering
the variable topography/bottom friction region.} \label{fig1}
\end{figure}

In particular, we shall be interested in the comparison of the slow
evolution of these two, initially identical, solitary waves in the
two different problems described above. The expected essential
difference in the evolution is due to the fact that the lead
solitary wave in the undular bore is generally not independent of
the remaining part of the bore and can exhibit features that cannot
be captured by a local perturbation analysis.  The well-known
example of such a behaviour, when a solitary wave is constrained by
the condition of being a part of a global nonlinear wave structure,
is provided by the  undular bore solution of the KdV-Burgers (KdV-B)
equation
\begin{equation}\label{kdvb}
u_t+6uu_x+u_{xxx}=\mu u_{xx}, \qquad \mu \ll 1 \, .
\end{equation}
Indeed, the undular bore solution of the KdV-B equation (\ref{kdvb})
is known to have a solitary wave at its leading edge (see Johnson
1970; Gurevich  $\&$ Pitaevskii 1987; Avilov, Krichever $\&$ Novikov
1987) and this solitary wave: (a) is asymptotically close to a
soliton solution of the unperturbed KdV equation; and (b) has the
amplitude, say $a_0$, that is constant in time. At the same time, it
is clear that if one takes an isolated KdV soliton of the same
amplitude $a_0$ as  initial data for the KdV-Burgers equation it
would damp  with time due to dissipation. The physical explanation
of such a drastic difference in the behaviour of an isolated soliton
and a lead solitary wave in the undular bore for the same weakly
dissipative KdV-B equation is that the action of weak dissipation on
an expanding undular bore is twofold: on the one hand, the
dissipation tends to decrease the amplitude of the wave locally but,
on the other hand, it  ``squeezes'' the undular bore so that the
interaction (i.e. momentum exchange) between separate solitons
within the bore becomes stronger than in the absence of dissipation
and this acts as the amplitude increasing factor. The
additional momentum is extracted from the upstream flow with a
greater depth (see Benjamin and Lighthill 1954).
As a result, in the case of the KdV-B equation, an
equilibrium non-zero value for the lead solitary wave amplitude in
the undular bore is established. Of course, for other types of
dissipation, a stationary value of the lead soliton amplitude would
not necessarily exist, but in general, due to the expected increase
of the soliton interactions near the leading edge, the amplitude of
the lead soliton of the undular bore would decay slower than that of
an isolated soliton.  Indeed, the presence here of  variable
topography as well can result in an additional ``nonlocal'' amplitude
growth.

While the problem (a) can be solved using traditional perturbation
analysis for a single solitary wave, which leads to an ordinary
differential equation along the solitary wave path (see Miles 1983a,b),
the undular bore evolution problem (b) requires a more
general approach which can be developed on the basis of Whitham's
modulation theory leading to a system of three nonlinear hyperbolic partial
differential equations of the first order. Since the Whitham method,
being  equivalent to a nonlinear multiple scale
perturbation procedure, contains the adiabatic theory of slow
evolution of a single solitary wave as a particular (albeit
singular) limit, it is instructive for the purposes of this paper to
treat both problems (a) and (b) using the general Whitham theory.

\section{Modulation equations}

The original Whitham method (Whitham 1965, 1974) was developed for
conservative constant-coefficient nonlinear dispersive equations and
is based on the averaging of appropriate conservation laws of the
original system over the period of a single-phase periodic
travelling wave solution. The resulting system of quasi-linear
equations describes the slow evolution of the modulations (i.e. of
the mean value, the wavenumber, the amplitude etc.) of the periodic
travelling wave. Here, that approach is extended to the perturbed
KdV equation (\ref{newvar}) following the general approach of
Kamchatnov (2004), which extends earlier results for certain
specific cases (see Gurevich and Pitaevskii (1987, 1991), Avilov,
Krichever and Novikov (1987) and Myint and Grimshaw (1995) for
instance).

We suppose that the evolution of the nonlinear wave is adiabatically
slow, that is, the wave can be locally represented as a solution of
the corresponding unperturbed KdV equation (i.e. (\ref{U}) with
zero on the right-hand side)
 with its parameters slowly varying with
space and time.  The one-phase periodic solution of the KdV
equation can be written in the form
\begin{equation}\label{cnoidal1}
    U(X,T)=\la_3-\la_1-\la_2-2(\la_3-\la_2)\sn^2(\sqrt{\la_3-\la_1}\,
    \theta,m)
\end{equation}
where $\sn(y,m)$ is the Jacobi elliptic sine function, $\la_1\le
\la_2 \le \la_3$ are  parameters and the phase variable $\theta$ and
the modulus $m$  are  given by
\begin{equation}\label{cnoidal2}
    \theta=X-VT,\quad V=-2(\la_1+\la_2+\la_3) \,,
\end{equation}
\begin{equation}\label{cnoidal3}
    m=\frac{\la_3-\la_2}{\la_3-\la_1} \,,
\end{equation}
\begin{equation}\label{cnoidal4}
 \hbox{and} \quad   L=\oint d\theta=\int_{\la_2}^{\la_3}\frac{d\mu}{\sqrt{-P(\mu)}}=\frac{2K(m)}
    {\sqrt{\la_3-\la_1}} \, ,
\end{equation}
where $K(m)$ is the complete elliptic integral of the first kind,  $L$ is
the ``wavelength'' along the $X$-axis (which is actually a retarded
time rather than a true spatial co-ordinate). Here we have used the
representation of the basic ordinary differential equation
 for the KdV travelling wave solution
(\ref{cnoidal1}) in the form (see Kamchatnov (2000) for a general
motivation behind this representation)
\begin{equation}\label{2-6}
    \frac{d\mu}{d\theta}=2\sqrt{-P(\mu)},
\end{equation}
where
\begin{equation}\label{2-5}
    \mu=\tfrac12(U+s_1), \quad s_1=\la_1+\la_2+\la_3
\end{equation}
and
\begin{equation}\label{2-7}
    P(\mu)=\prod_{i=1}^3(\mu-\la_i)=\mu^3-s_1\mu^2+s_2\mu-s_3,
\end{equation}
that is the solution (\ref{cnoidal1}) is parameterized by the zeroes
$\la_1,\,\la_2,\,\la_3$ of the polynomial $P(\mu)$.

In a  modulated wave, the parameters $\la_1,\,\la_2,\,\la_3$ are
allowed to be slow functions of $X$ and $T$, and their
evolution is governed by the Whitham equations. For the
unperturbed KdV equation, the evolution of the modulation
parameters is  due to a spatial non-uniformity of the initial
distributions for $\la_j , j=1,2,3$ and the typical spatio-temporal scale
of the modulation variations is determined by the scale of the
initial data.

 In the case of the perturbed KdV
equation (\ref{U}), the evolution of the parameters
$\la_1,\,\la_2,\,\la_3$ is caused not only by their initial spatial
non-uniformity, but also by the action of the weak perturbation, so
that, generally, at least two independent spatio-temporal scales {
\it for the modulations} can be involved.  However, at this point we
shall not introduce any scale separation within the modulation
theory and derive general perturbed Whitham equations assuming that
the typical values of $F(T)$ and $G(T)$ are $\mathcal{O}(\prt
\la_j/\prt T , \prt \la_j /\prt X)$ within the modulation theory.

It is instructive to first introduce the Whitham equations for the
perturbed KdV equation (\ref{U}) using the traditional approach
of averaging the (perturbed) conservation laws. To this end, we
introduce the averaging over the period (\ref{cnoidal4}) of the cnoidal
wave (\ref{cnoidal1}) by
\begin{equation}\label{average}
    \langle\mathfrak{F}\rangle=\frac1L\oint\mathfrak{F}d\theta =
    \frac1L \int_{\la_2}^{\la_3}\frac{\mathfrak{F}d\mu}{\sqrt{-P(\mu)}}.
\end{equation}
In particular,
\begin{equation}\label{avU}
\langle U\rangle=2\langle\mu\rangle-s_1=
    2(\la_3-\la_1)\frac{E(m)}{K(m)}+\la_1-\la_2-\la_3,
\end{equation}
\begin{equation}\label{avU2}
\langle U^2 \rangle =
8[-\frac{s_1}{6}(\la_3-\la_1)\frac{E(m)}{K(m)} -
\frac{1}{3}s_1\la_1+\frac{1}{6}(\la_1^2 - \la_2\la_3)] +s_1^2 \,,
\end{equation}
where $E(m)$ is the complete elliptic integral of the second kind.
Now, one represents the KdV equation (\ref{U}) in the form of
the perturbed conservation laws
\begin{equation}\label{cons}
\frac{\partial P_j}{\partial T}+ \frac{\partial Q_j}{\partial X}=
 R_j  \, , \quad j=1,2,3\, , \quad R_j \ll 1 \, ,
\end{equation}
where $P_j$ and $Q_j$ are the standard expressions for the conserved
densities (Kruskal integrals) and ``fluxes''  of the unperturbed KdV
equation.  Just as in the Whitham (1965) theory for unperturbed
dispersive systems, the number of  conservation laws required is
equal to the number of free parameters in the travelling wave
solution, which is three in the present case.
Next, one applies the averaging (\ref{average}) to the system
(\ref{cons}) to obtain (see Dubrovin and Novikov 1989)
\begin{equation}\label{avcons}
\frac{\partial \langle P_j\rangle}{\partial T}+ \frac{\partial
\langle Q_j \rangle}{\partial X}= \langle R_j \rangle \, , \quad
j=1,2,3\, .
\end{equation}
The system (\ref{avcons}) describes slow evolution of the parameters
$\la_j$ in the cnoidal wave solution (\ref{cnoidal1}).

Along with these derived perturbed conservative form of the Whitham
equations, we introduce the wave conservation law which is a general
condition for the existence of slowly modulated single-phase
travelling wave solutions (\ref{cnoidal1}) (see for instance Whitham
1974) and must be consistent with the modulation system
(\ref{avcons}). This conservation law has the form
\begin{equation}\label{wc}
\frac{\partial k}{\partial T}+\frac{\partial \omega}{\partial X}=0
\, ,
\end{equation}
\begin{equation}\label{kV}
\hbox{where} \quad k=\frac{2\pi}{L} \, , \qquad \omega=kV
\end{equation}
are the ``wavenumber'' and the ``frequency'' respectively (we have
put quotation marks here because the actual wavenumber and frequency
related to the physical variables $x,t$ are different
quantities from those in (\ref{kV}),  but are related through the
transformations (\ref{1-3}, \ref{newvar}) ). The wave
conservation law (\ref{wc}) can be introduced instead of any of
three inhomogeneous averaged conservation laws comprising the
Whitham system (\ref{avcons}).

It is known that the Whitham system for the homogeneous
constant-coefficient KdV equation can be represented in diagonal
(Riemann) form (Whitham 1965, 1974) by an appropriate choice of the
three parameters characterising the periodic travelling wave
solution. In fact, in our solution (\ref{U}) the parameters
$\la_j$ have already been chosen so that they coincide with the
Riemann invariants of the unperturbed KdV modulation system.
Introducing them explicitly into the perturbed system (\ref{avcons})
we obtain (see Kamchatnov 2004)
\begin{equation}\label{pertmod}
    \frac{\prt\la_i}{\prt T}+v_i\frac{\prt\la_i}{\prt X}=
    \frac{L}{\prt L/\prt\la_i}\cdot
    \frac{\langle(2\la_i-s_1-U)R\rangle}{4\prod_{j\neq
    i}(\la_i-\la_j)},\quad i=1,2,3,
\end{equation}
where $R$  is the  perturbation term on the right-hand side of the KdV
equation (\ref{U}) and
\begin{equation}\label{chvel}
    v_i=-2\sum\la_i+\frac{2L}{\prt L/\prt\la_i},\quad i=1,2,3,
\end{equation}
are the Whitham characteristic velocities corresponding to the
unperturbed KdV equation.

It should be noted that the straightforward realisation of the above
lucid general algorithm for obtaining  perturbed modulation system
in diagonal form is quite a laborious task. In fact, to derive
system (\ref{pertmod}), the so-called finite-gap integration method
incorporating the integrable structure of the unperturbed KdV
equation has been used.  The modulation system (\ref{pertmod}) in a
more particular form corresponding to specific choices of the
perturbation term was obtained by Myint and Grimshaw (1995) using a
multiple-scale perturbation expansion.  In that latter setting, the
wave conservation law (\ref{wc}) is an inherent part of the
construction, while in the averaging approach used here, it can be
obtained as a consequence of the system (\ref{pertmod}).

To obtain an explicit representation of the Whitham equations for
the present case of equation (\ref{U}), we  must substitute the
perturbation $R$ from the right-hand side of (\ref{U}) and perform
the integration (\ref{average}) with $U$ given by (\ref{cnoidal1}).
From now on, we are going to consider only the flows where $U \ge 0$
so that the perturbation term assumes the form
\begin{equation}\label{pert}
R(U)=G(T)U-F(T)U^2 \, .
\end{equation}

Substituting (\ref{pert}) into (\ref{pertmod}) we obtain, after some
detailed calculations (see Appendix), the perturbed
Whitham system in the form
\begin{equation}\label{5-2}
    \frac{\prt\la_i}{\prt T}+v_i\frac{\prt\la_i}{\prt X}=
    \rho_i=C_i[F(T)A_i-G(T)B_i], \quad i=1,2,3
\end{equation}
\begin{equation}\label{5-3}
 \hbox{where} \quad    C_1=\frac1{E},\quad C_2=\frac1{E-(1-m)K},\quad C_3=\frac1{E-K};
\end{equation}
\begin{equation}\label{5-4}
    \begin{split}
    A_1&=\frac13(5\la_1-\la_2-\la_3)E+\frac23(\la_2-\la_1)K,\\
    A_2&=\frac13(5\la_2-\la_1-\la_3)E-(\la_2-\la_1)
    \left(\frac13+\frac{\la_2}{\la_3-\la_1}\right)K,\\
    A_3&=\frac13(5\la_3-\la_1-\la_2)E-\left[\la_3+\frac13(\la_2-\la_1)\right]K;
    \end{split}
\end{equation}
\begin{equation}\label{5-5}
    \begin{split}
    B_1&=\frac1{15}(-27\la_1^2-7\la_2^2-7\la_3^2+2\la_1\la_2+2\la_1\la_3+22\la_2\la_3)E\\
    &-\frac4{15}(\la_2-\la_1)(3\la_1+\la_2+\la_3)K,\\
    B_2&=\frac1{15}(-7\la_1^2-27\la_2^2-7\la_3^2+2\la_1\la_2+22\la_1\la_3+2\la_2\la_3)E\\
    &+\frac1{15}\frac{\la_2-\la_1}{\la_3-\la_1}(7\la_1^2+15\la_2^2+11\la_3^2-6\la_1\la_2
    -18\la_1\la_3+6\la_2\la_3)K,\\
    B_3&=\frac1{15}(-7\la_1^2-7\la_2^2-27\la_3^2+22\la_1\la_2+2\la_1\la_3+2\la_2\la_3)E\\
    &+\frac1{15}(7\la_1^2+11\la_2^2+15\la_3^2-18\la_1\la_2
    -6\la_1\la_3+6\la_2\la_3)K;
    \end{split}
\end{equation}
and the characteristic velocities are:
\begin{equation}\label{5-6}
    \begin{split}
    v_1&=-2\sum\la_i+\frac{4(\la_3-\la_1)(1-m)K}E,\\
    v_2&=-2\sum\la_i-\frac{4(\la_3-\la_2)(1-m)K}{E-(1-m)K},\\
    v_3&=-2\sum\la_i+\frac{4(\la_3-\la_2)K}{E-K}.
    \end{split}
\end{equation}

The equations (\ref{5-2}) -- (\ref{5-6}) provide a general setting for
studying the nonlinear modulated wave evolution over variable
topography with bottom friction. In the absence of the perturbation
terms (i.e. when $F(T) \equiv 0$, $G(T) \equiv 0$), the system
(\ref{5-2}), (\ref{5-6}) indeed coincides with the
original Whitham equations (Whitham 1965)  for the integrable KdV
dynamics. In that case the variables $\la_1,\,\la_2,\,\la_3$ become
Riemann invariants, so in this general (perturbed) case we shall call
them Riemann variables.

It is important to study the structure of the perturbed Whitham
equations (\ref{5-2}) -- (\ref{5-6}) in two limiting cases when the
underlying cnoidal wave degenerates into (i) a small-amplitude
sinusoidal wave (linear limit), when $\la_2=\la_3$ $(m=0)$, and (ii)
into a solitary wave when $\la_2=\la_1$ $(m=1)$. Since in both these
limits the oscillations do not contribute to the mean flow (they are
infinitely small in the linear limit and the distance between them
becomes infinitely long in the solitary wave limit) one should
expect that in both cases one of the Whitham equations will
transform into the dispersionless limit of the original perturbed
KdV equation (\ref{U}) i.e.
\begin{equation}\label{6-1}
    U_T+6UU_X=F(T)U-G(T)U^2,
\end{equation}
Indeed, using formulae (\ref{5-2}) -- (\ref{5-6}) we obtain for
$m=0$:
\begin{equation}\label{6-3}
\begin{split}
    &\la_2=\la_3 \, , \\
    &\frac{\prt\la_1}{\prt T}-6\la_1\frac{\prt\la_1}{\prt X}=\la_1F+\la_1^2G,\\
    &\frac{\prt\la_3}{\prt T}+(6\la_1-12\la_3)\frac{\prt\la_3}{\prt X}=
    \la_1F+\la_1^2G \, .
    \end{split}
\end{equation}
Similarly, for $m=1$, one has
\begin{equation}\label{6-2}
\begin{split}
    &\la_2=\la_1 \, , \\
    &\frac{\prt\la_1}{\prt T}-(4\la_1+2\la_3)\frac{\prt\la_1}{\prt X}=
    \frac{1}3(4\la_1-\la_3)F+\frac{1}{15}(7\la_3^2-24\la_1\la_3+32\la_1^2)G,\\
    &\frac{\prt\la_3}{\prt T}-6\la_3\frac{\prt\la_3}{\prt
    X}=\la_3F+\la_3^2G\,.
     \end{split}
\end{equation}
We see that, in both cases, one of the Riemann variables (taken with
inverted sign) coincides with the solution of the dispersionless
equation (\ref{6-1}) (recall that in the derivation of the Whitham
equations we assumed $U\geq0$ everywhere), namely $U=\langle U
\rangle =-\la_1$ when $\la_2=\la_3$ $(m=0)$ and $U= \langle U
\rangle =-\la_3$ when $\la_2=\la_1$ $(m=1)$.

To conclude this section, we present expressions for the
physical wave parameters such as the surface elevation wave
amplitude $a$, mean elevation $\langle A \rangle$ speed and wavenumber
 in terms of  the modulation solution $\la_j(X,T)$. Using (\ref{newvar}) and
(\ref{cnoidal1}) we obtain for the wave amplitude (peak to trough)
and the mean elevation
\begin{equation}\label{ampl}
a=\frac{4h^2}{3g}(\la_3-\la_2) \, , \qquad \langle A \rangle =
\frac{2h^2}{3g}\langle U \rangle \, ,
\end{equation}
where the dependence of $\langle U \rangle$  on $\la_j(X,T), \ j=1,2,3$
is given by (\ref{avU}) and $X=X(x,t)$, $T=T(x,t)$ by (\ref{1-3},
\ref{newvar}). In order to obtain the physical wavenumber $
\kappa$ and the frequency $\Omega$ we first note that the
phase function $\theta (X,T)$ defined in (\ref{cnoidal2}) is replaced by
a more general expression defined so that $k = \theta_X $
and $kV = - \theta_T $ are the ``wavenumber'' and
``frequency'' in the $X-T$ coordinate system.   Then we define the
physical phase function $\Theta (x,t) = \theta (X, T)$  so that we get
\begin{equation}\label{kO}
\kappa=\Theta_x\, , \qquad \Omega=-\Theta_t \, .
\end{equation}
It now follows that
\begin{equation}\label{Om}
\kappa = \frac{k}{c}(1- \frac{hV}{6g}) \,, \quad \Omega = k \,,
\quad \hbox{and} \quad \frac{\Omega }{\kappa } =
\frac{c}{1-hV/6g} \,.
\end{equation}\\
 Note that the physical frequency is the
``wavenumber'' in the $X-T$ coordinate system, and that the physical
phase speed is $\Omega/\kappa  $. Since the validity of the KdV
model (\ref{1-1}) requires {\it inter alia} that the wave be
right-going it follows from this expression that the modulation
solution remains valid only when $hV< 6g$. Of course, the validity
of (\ref{1-1}) also requires that the amplitude remains small, and
this would normally also ensure that $V$ remains small.

\section{Modulation solution in the solitary wave limit}
In this section, we shall integrate the  perturbed modulation system
(\ref{5-2})  along the multiple characteristic corresponding to the
merging of two Riemann variables $\la_2$ and $\la_1$. As we shall
see later, this characteristic specifies the motion of the leading
edge of the shoaling undular bore in the case when the perturbations
due to variable topography and bottom friction can be considered as
small compared with the existing spatial modulations within the
bore. At the same time, as the case $\la_2=\la_1$ ( i.e. $m=1$)
corresponds to the solitary wave limit in the travelling wave
solution (\ref{cnoidal1}), our results here are expected to be
consistent with the results from the traditional perturbation approach
to the adiabatic variation of a solitary wave due to topography and
bottom friction (see Miles 1983a,b).

In the limit  $m\to1$ the periodic solution (\ref{cnoidal1}) of the
KdV equation goes over to its solitary wave solution
\begin{equation}\label{7-1}
    U(X,T)=U_0\hbox{sech}^2[\sqrt{\la_3-\la_1}(X-V_sT)]-\la_3,
\end{equation}
where
\begin{equation}\label{uv}
U_0=2(\la_3-\la_1)\, , \quad  V_s=-(4\la_1+2\la_3)\,
\end{equation}
are the solitary wave amplitude and ``velocity'' respectively. The
solution (\ref{7-1}) depends on two parameters $\la_1$ and $\la_3$
whose adiabatic slow evolution is governed by the reduced modulation
system (\ref{6-2}). It is important that the second equation in this
system is decoupled from the first one. Hence, evolution of the
pedestal $-\la_3$ on which the solitary wave rides, can be found
from the solution of this dispersionless equation by the method of
characteristics. When $\la_3(X,T)$ is known, evolution of the
parameter $\la_1$ can be found from the solution of the first
equation (\ref{6-2}). As a result, we arrive at a complete
description of adiabatic slow evolution of the solitary wave
parameters taking account of its interaction with the (given)
pedestal.

However, it is important to note here that while this description of
the adiabatic evolution of a solitary wave is complete as far as the
solitary wave itself is concerned, it fails to describe the
evolution of a trailing shelf, which is needed to conserve total
``mass'' (see, for instance, Johnson 1973b, Grimshaw 1979 or
Grimshaw 2006). This trailing shelf has a very small amplitude, but
a very large length scale, and hence can carry the same order of
``mass'' as the solitary wave. But note that the ``momentum'' of the
trailing shelf is much smaller than that of the solitary wave, whose
adiabatic deformation is in fact governed to leading order by
conservation of ``momentum'', or more precisely, by conservation of
wave action flux (strictly speaking, conservation only in the
absence of friction).

The situation simplifies if the solitary wave propagates into a
region of still water so that there is no pedestal ahead of the wave, that is
$\la_3 = 0$ in $X>\tau(T)$.   But then, since $\la_3=0$ is an exact solution of
the degenerate Whitham system  (\ref{6-2}) for this solitary wave configuration,
we can put $\la_3=0$ both in the solitary wave solution,
\begin{equation}\label{7-2}
    U(X,T)=-2\la_1\hbox{sech}^2[\sqrt{-\la_1}\,(X-V_sT)],
    \quad V_s=-4\la_1,
\end{equation}
and in  equation (\ref{6-2}) for the parameter $\la_1$ to obtain,
\begin{equation}\label{8-1}
    \frac{\prt\la_1}{\prt T}-4\la_1\frac{\prt\la_1}{\prt X}
    =\frac43F\la_1+\frac{32}{15}G\la_1^2 \, ,
\end{equation}
As we see, the solitary wave moves with the instant velocity
\begin{equation}\label{8-2}
    \frac{dX}{dT}=-4\la_1,
\end{equation}
and the parameter $\la_1$ changes with $T$ along the solitary wave
trajectory according to the ordinary differential equation
\begin{equation}\label{8-3}
    \frac{d\la_1}{dT}=
    \frac43F(T)\la_1+\frac{32}{15}G(T)\la_1^2.
\end{equation}
It can be shown that equation (\ref{8-3}) is consistent with the
equation for the solitary wave half-width $\gamma=\sqrt{-\la_1}$
obtained by the traditional perturbation approach (see Grimshaw
(1979) for instance).

Next, we re-write equation (\ref{8-3}) in terms the original
independent $x$-variable.   For that, we find from (\ref{newvar}),
that
\begin{equation}\label{dT}
dT=(h^{1/2}/6g^{3/2})dx
\end{equation}
\begin{equation}\label{8-4}
\hbox{and} \quad     F=-\frac{27}2\left(\frac{g}h\right)^{3/2}\frac{dh}{dx},\quad
    G=4C_D\left(\frac{g}h\right)^{1/2}.
\end{equation}
Then substituting these expressions into (\ref{8-3}) yields
the equation
\begin{equation}\label{8-5}
    \frac{d\la_1}{dx}=-3\frac1h\frac{dh}{dx}\la_1+\frac{64}{45}
    \frac{C_D}{g}\la_1^2
\end{equation}
which can be easily integrated to give
\begin{equation}\label{8-6}
    \frac1{\la_1}=h^3\left(-C_0-\frac{64}{45}\frac{C_D}g
    \int_0^x\frac{dx}{h^3}\right),
\end{equation}
where $C_0$ is an integration constant and $x=0$ is a reference
point where $h=h_0$. According to (\ref{7-2}), $U_0=-2\la_1$ is the
amplitude of the soliton expressed in terms of variable $U(X,T)$.
Returning to the original surface displacement $A(x,t)$ by means of
(\ref{newvar}) and denoting $C_0=4/(3ga_0h_0)$, we find the
dependence of the surface elevation soliton amplitude
$a=(2h^2/3g)U_0$ on $x$ in the form
\begin{equation}\label{8-7}
    a=a_0\left(\frac{h_0}h\right)\left[1+\frac{16}{15}C_Da_0h_0
    \int_0^x\frac{dx}{h^3}\right]^{-1},
\end{equation}
where $a_0$ is the solitary wave amplitude at $x=0$. We note that
for $C_D=0$ this reduces to the classical  Boussinesq (1872) result
$a \sim h^{-1}$, while for $h=h_0 $ it reduces to the well-known
algebraic decay law $a \sim 1/(1+\hbox{constant} \ x) $ due to Chezy
friction. Miles (1983a,b) obtained this expression for a linear
depth variation, although we note that there is a factor of $2$
difference from (\ref{8-7}) (in Miles (1983a,b) the factor
$16C_D/15$ is $8C_D/15$). The trajectory of the soliton can be now
found from (\ref{8-2}) and (\ref{8-6}):
\begin{equation}\label{8-8}
    X=\int_0^x\frac{dx}{\sqrt{gh}}-t=\frac{a_0h_0}{2\sqrt{g}}
    \int_0^x dx' h^{-5/2}(x')\left[1+\frac{16}{15}C_Da_0h_0
    \int_0^{x'}\frac{dx}{h^3(x)}\right]^{-1}.
\end{equation}
This expression determines implicitly the dependence of $x$ on $t$
along the solitary wave  path and provides the desired equation for
the multiple characteristic of the modulation system for the case
$m=1$.

It is instructive to derive an explicit expression for the solitary
wave speed by computing the derivative $dx/dt$ from (\ref{8-8}), or
more simply, directly from (\ref{Om}),
\begin{equation}\label{v}
v_s =\frac{dx}{dt}=  \frac{c}{1-{a}/{2h}}\, .
\end{equation}
The formula (\ref{v}) yields the restriction for the relative
amplitude $ \gamma =a/h <2$ which is clearly beyond the
applicability of the  KdV approximation  (wave breaking occurs
already at $\gamma=0.7$ (see Whitham 1974)). In the frictionless
case ($C_D = 0$) equation (\ref{8-7}) gives $a/h =a_0 h_ 0/h^2 $,
and so the expression (\ref{v}) for the speed must fail as $h \to
0$. It is interesting to note that this failure of the KdV model as
$h \to 0$ due to appearance of infinite (and further negative!)
solitary wave speeds is not apparent from the expression (\ref{8-7})
for the solitary wave amplitude, and the implication is that the
model cannot be continued as $h \to 0$. Curiously this restriction
of the KdV model seems never to have been noticed before in spite of
numerous works on this subject.  Note that taking account of bottom
friction leads to a more complicated formula for the solitary wave
speed as a function of $h$ but the qualitative result remains the
same.

It is straightforward to show from (\ref{8-5}) or (\ref{8-7}) that
\begin{equation}\label{8-9}
\frac{a_x }{a} = -\frac{h_x }{h} -\frac{16}{15}\frac{C_D a_0 h_0 }{h^3}
\left[1+\frac{16}{15}C_Da_0h_0 \int_0^x\frac{dx}{h^3}\right]^{-1} \,.
\end{equation}
It follows immediately that for a wave advancing into increasing depth
($h_x > 0$), the amplitude decreases due to a combination of increasing depth
and bottom friction. However, for a wave advancing into decreasing depth,
there is a tendency to increase the amplitude due to the depth decrease, but to decrease the
amplitude due to bottom friction. Hence whether or not the amplitude increases
is determined by which of these effects is larger, and this in turn is determined
by the slope, the depth,  and the consolidated drag parameter $C_D a_0/h_0 $.

To illustrate, let us consider the bottom topography in the form
\begin{equation}\label{9-1}
  h(x) = h_{0}^{1-\al }(h_0 -\delta x)^{\al } \,, \quad \al > 0 \,,
\end{equation}
which satisfies the  condition $h(0)=h_0$; the parameter $\delta$
characterizes the slope of the bottom. In this case
the formula (\ref{8-7})  becomes\\
\begin{equation}\label{9-2}
 a=a_0\left(\frac{h_0}h\right)\left[1+\frac{16}{15}\frac{C_D a_0 }
 {\delta (3\al -1) h_0 }\left\{\left(\frac{h_0}{h}\right)^{(3\al -1)/\alpha }-1\right\}
  \right]^{-1} \,
\end{equation}
if $\alpha\ne 1/3$. One can see now that if $\al<1/3$, then the
bottom friction term is relatively unimportant due to the smallness
of $C_D $. Of course, for this case we again recover the Boussinesq
result, now slightly modified,
\begin{equation}\label{9-3}
    a  \approx  a_0 \frac{h_0}{h} \left[1+\frac{16}{15}\frac{C_D a_0 }
    {\delta (1-3\al ) h^{2}_0  }\right]^{-1}  \,, \quad 0<\al<\frac13,\quad h \ll h_0.
\end{equation}
Of course, this result  is impractical in the KdV context  as the
KdV approximation used here requires the ratio $a/h$ to remain
small.

If $\al>1/3$  now obtain asymptotic formula
\begin{equation}\label{9-4}
    a \approx   \frac{15(3\al-1)\delta}{16C_D}h_0\left(\frac{h_0}h\right)^{\frac1{\al}-2},
    \quad  h\ll h_0 \,,
\end{equation}
which is independent of the initial amplitude $a_0$.  This
expression is consistent with the small-amplitude KdV
approximation as long as $(3\al -1) \delta /C_D$ is order unity.
Simple inspection of  (\ref{9-4}) shows that  the solitary wave amplitude
\begin{itemize}
    \item increases as $h\to 0$ if $\tfrac13<\al<\tfrac12$,
    \item is constant as $h\to 0$ if $\al=\tfrac12$,
    \item decreases as $h\to0$ if $\al>\tfrac12$.
\end{itemize}
Thus for $1/3 <\al <1/2$,  as for the case $\al < 1/3$, the
amplitude will increase as the depth decreases, in spite of the
presence of (sufficiently small) friction. However, for $\al > 1/3
$, even although there is usually some initial growth in the
amplitude, eventually even small bottom friction will take effect
and the amplitude decreases to zero.  We note that if $\alpha =1/3$
then the integral $\int^x_0h^{-3}dx $ in (\ref{8-7}) diverges
logarithmically as $h \to 0 $, which just slightly modifies the
result (\ref{9-4}) for $h \ll h_0$ and implies growth of the
amplitude $\propto \ln h/ h$ as $h \to 0$.

Of particular interest is the case  $\al=1$. In that case formula
(\ref{9-2}) becomes
\begin{equation}\label{miles}
 a=a_0\left(\frac{h_0}h\right)\left[1+\frac{8}{15}\frac{C_D a_0 }
 {\delta  h_0 }\left\{\left(\frac{h_0}{h}\right)^{2 }-1\right\}
  \right]^{-1} \,.
\end{equation}
\begin{equation}\label{miles1}
\hbox{and} \quad a \approx \frac{15}{8}\frac{\delta}{C_D}h\, , \qquad h \ll h_0
\end{equation}
These expressions  (\ref{miles}, \ref{miles1}) were obtained by
Miles (1983a,b) using wave energy conservation (as above, note,
however, that in Miles (1983a,b)
 the numerical coefficient is $15/4$ rather than $15/8$).
Thus, these results obtained from the Whitham theory  are indeed
consistent, at the leading order, with the traditional perturbation approach for a
slowly-varying solitary wave.

\section{Adiabatic deformation of a cnoidal wave}

Next we consider a modulated cnoidal wave (\ref{cnoidal1}) in the
special case when the modulation does not depend on $X$. While this
case is, strictly speaking, impractical as it assumes there is an
infinitely long wavetrain, it can nevertheless provide some useful
insights into the  qualitative effects of gradual slope and friction
on undular bores which are locally represented as cnoidal waves. In
the absence of friction, the slow dependence of the cnoidal wave
parameters on $T$ was obtained by Ostrovsky $\&$ Pelinovsky (1970,
1975) and Miles (1979) (see also Grimshaw 2006),
 assuming that the surface displacement had a zero mean
 (i.e. $\langle U \rangle=0$),
while, the effects of friction were taken into account by Miles
(1983b) using the same zero-mean displacement assumption.
However, this assumption is inconsistent with our aim to study
undular bores where the value of  $\langle U \rangle $ is
essentially nonzero. Hence, we need to develop a more general theory
enabling us to take into account variations in all the parameters in
the cnoidal wave. Such a general setting is provided by the
modulation system (\ref{5-2}).

Thus we consider  the case when the Riemann variables in (\ref{5-2})
do not depend on the variable $X$ so that the general Whitham
equations become ordinary differential equations in $T$, which can
be conveniently reformulated in terms of the original spatial
$x$-coordinate using the relationship (\ref{dT}):
\begin{equation}\label{14-1}
    \frac{d\la_i}{dx}=C_i\left[-\frac94\frac1h\frac{dh}{dx}A_i-\frac{2C_D}{3g}B_i
    \right],\qquad i=1,2,3,
\end{equation}
where all variables are defined above in section 3 (see \ref{5-3},
\ref{5-4}, \ref{5-5}). This system can be readily solved
numerically. But it is instructive, however, to first indicate some
general properties of the solution.

First, the solution to the system (\ref{14-1}) must have the
property of conservation of  ``wavelength'' $L$ (or ``wavenumber''
k=$2\pi/L$)
\begin{equation}\label{14-2}
    L=\frac{2K(m)}{\sqrt{\la_3-\la_1}}=\mathrm{constant}
\end{equation}
Indeed, the wave conservation law (\ref{wc}) in absence of
$X$-dependence assumes the form
\begin{equation}\label{wc1}
\frac{\partial k}{\partial T}=0 \, ,
\end{equation}
which yields (\ref{14-2}). Thus the system of three equations (\ref{14-1})
can be reduced to two equations.

Next, applying Whitham averaging directly to (\ref{U}) yields
\begin{equation}\label{14-3}
\frac{dM}{dx} = -\frac94\frac1h\frac{dh}{dx}M-\frac{2C_D}{3g}\tilde{P} \,, \quad
M=\langle U  \rangle \,, \quad \tilde{P}  = \langle|U|U  \rangle \,.
\end{equation}
\begin{equation}\label{14-4}
\frac{dP}{dx} =
-\frac92\frac1h\frac{dh}{dx}P-\frac{4C_D}{3g}\tilde{Q} \,, \quad
P=\langle U^2  \rangle \,, \quad \tilde{Q}  = \langle|U|^3  \rangle
\,.
\end{equation}
The equation set (\ref{14-2}), (\ref{14-3}), (\ref{14-4}) comprise a
closed modulation system for three independent modulation
parameters, say $M$, $\tilde {P}$ and $m$. While this system is not
as convenient for further analysis as the system (\ref{5-2}) in
Riemann variables, it does not have a restriction $U>0$ inherent in
(\ref{5-2}), and allows for some straightforward inferences
regarding the possible existence of modulation solutions with zero
mean elevation, that is with $M=0$. Indeed, one can see that the
solution with the zero mean is actually not generally permissible
when $C_D \ne 0$, a situation overlooked in Miles (1983b). Indeed,
$M=0$ immediately then implies that  $\tilde{P} =0$ by (\ref{14-3}).
But then due to (\ref{14-2}) we have all three modulation parameters
fixed which is clearly inconsistent with the remaining equation
(\ref{14-4}) (except for the trivial case $M=0$, $P=0$, $\tilde
Q=0$).  However, in the absence of friction, when $C_D=0$, equation
(\ref{14-3}) uncouples and permits a nontrivial solution  with a
zero mean. In general, when $C_D = 0$ equations (\ref{14-3}),
(\ref{14-4}) can be easily integrated to give
\begin{equation}\label{14-5}
    d= Mh^{9/4}=\hbox{constant}; \quad \sigma = Ph^{9/2}=
    \hbox{constant}.
    \end{equation}
Then, using (\ref{avU}, \ref{avU2}, \ref{14-2}) one readily gets the
formula for the variation of the modulus $m$, and hence of all the
other wave parameters, as a function of $h$
\begin{equation}\label{14-6}
  K^2 [2(2-m)EK -3E^2-(1-m)K^2] =
  \left(\frac{4}{3}\right)^5\frac{(\sigma - d^2)L^4}{h^{9/2}} \,.
\end{equation}
Formula (\ref{14-6}) generalises to the case $M \ne 0$ (i.e. $d \ne
0$) the expressions of Ostrovsky $\&$ Pelinovsky (1970, 1975), Miles
(1979) and Grimshaw (2006) (note that in Grimshaw (2006) the zero
mean restriction in actually not necessary). We note here  that,
again with $C_D = 0$, equation (\ref{kdvB}) implies conservation of
$\langle B \rangle$ and $\langle B^2\rangle$ (the averaged wave
action flux), which, together with (\ref{14-2}), also yield
(\ref{14-6}).

 The physical frequency $\Omega$ and  wavenumber
$\kappa$ in the modulated periodic wave under study are given by the
formula (\ref{Om}), and we recall here that $k=2\pi/L$ is constant
(see (\ref{14-2})). As discussed before at the end of Section 3 we
must require that the phase speed stays positive as the wave
evolves, and here that requires that the physical wavenumber $\kappa
> 0$. Since $a/h$ (and hence $hV/6g$)  is supposed to be small
within the range of applicability of the KdV equation (\ref{1-2})
the expression (\ref{Om}) implies the behaviour $\kappa \simeq
{\Omega}/{\sqrt{gh}} $ which of course agrees with the well known
result for linear waves on a sloping beach (see Johnson 1997 for
instance). This effect will be slightly attenuated for the nonlinear
cnoidal wave, since $Vh/6g
> 0$, but the overall effect will be a ``squeezing'' of the cnoidal wave,
a result important for our further study of undular bores. Next we
study numerically the combined effect of slope and friction on a
cnoidal wave.
\begin{figure}[bt]
\centerline{\includegraphics[width=8cm,height=6cm,clip]{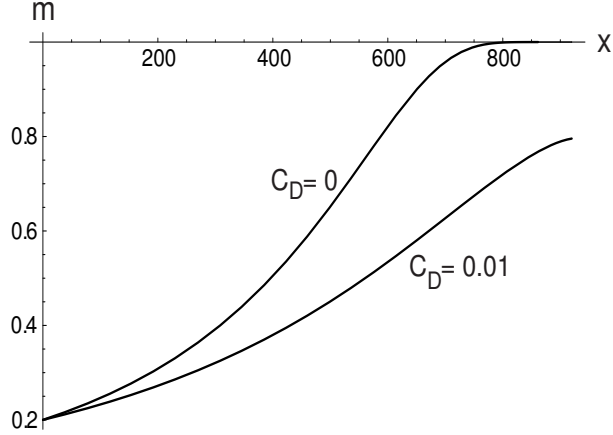}}
\caption{Dependence of the modulus $m$ on the physical space
coordinate $x$ in the cases without and with bottom friction in the
$X$-independent modulation solution.} \label{fig2}
\end{figure}

\begin{figure}[bt]
\centerline{
\includegraphics[width=8cm,height=6cm,clip]{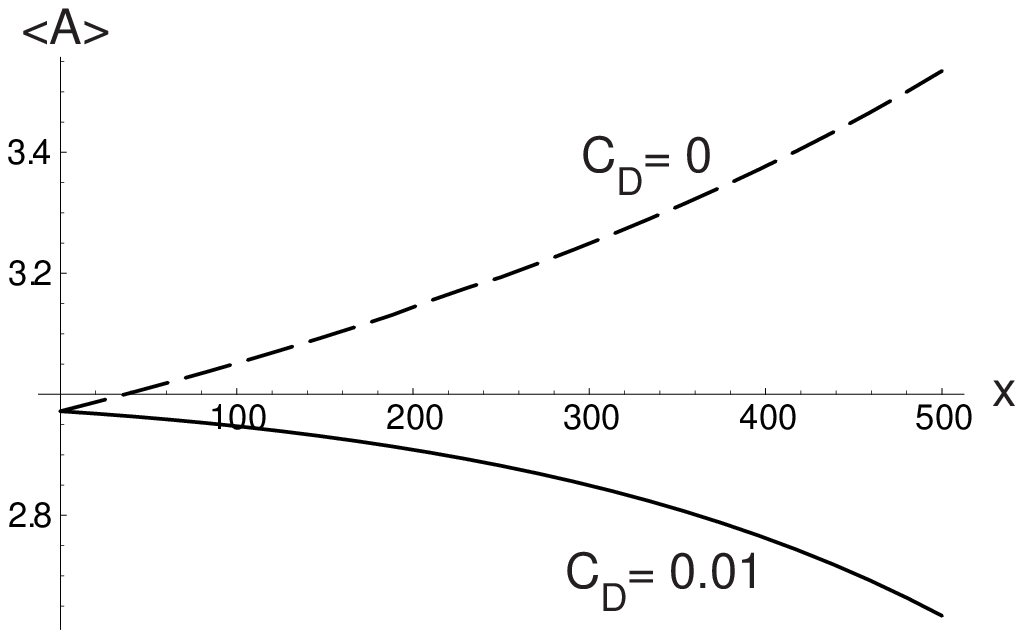}
\includegraphics[width=8cm,height=6cm,clip]{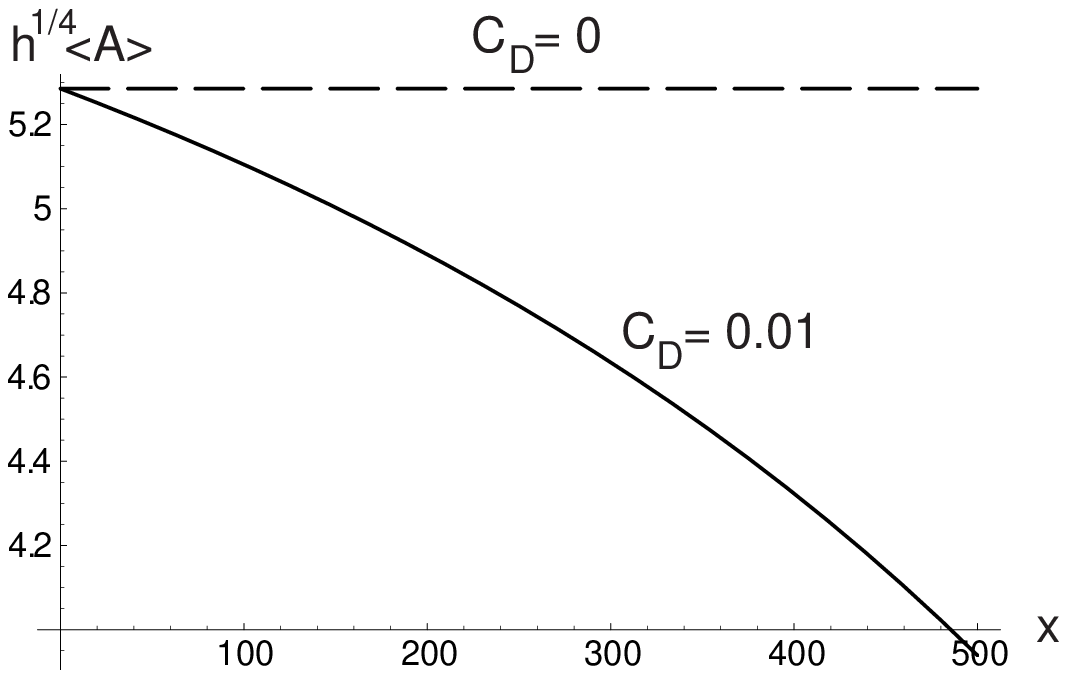}}
\vspace{0.3 true cm} \caption{Left: Dependence of the mean value
$\langle A \rangle $ in the $X$-independent modulation solution on the
physical space coordinate $x$  without (dashed line) and with (solid
line) bottom friction; Right: Same but multiplied by the Green's law
factor, $h^{1/4} $ } \label{fig3}
\end{figure}

\begin{figure}[bt]
\centerline{\includegraphics[width=8cm,height=6cm,clip]{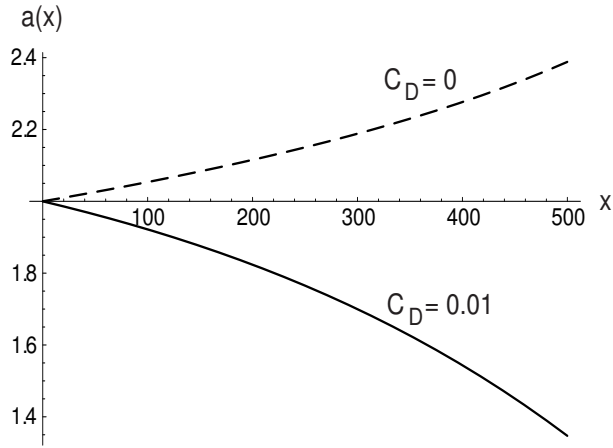}}
\caption{Dependence of the surface elevation amplitude $a$ on the space
coordinate $x$. Dashed line corresponds to the frictionless case and
solid line to the case with bottom friction.} \label{fig4}
\end{figure}
As we have shown, in the presence of Chezy friction $M \ne 0$, and
we have also assumed  that $U>0$, which is necessary when we come to
study undular bores. Now we use  the stationary modulation system
(\ref{14-1}) in Riemann variables, which was derived using this
assumption. We solve the coupled ordinary differential equation
system (\ref{14-1}) for the case of a linear slope
\begin{equation}\label{15-1}
    h(x)=h_0-\delta x
\end{equation}
with $h_0=10$, $\delta=0.01$,  and with the initial conditions
\begin{equation}\label{15-2}
    \la_1=-0.441,\quad \la_2=0.147,\quad \la_3= 0.294 \quad\mathrm{at}\quad x=0,
\end{equation}
which corresponds to  a nearly harmonic wave with $m=0.2$,
$a/h_0=0.2$, $\langle A \rangle /h_0 \approx 0.3$ at $x=0$ (see
(\ref{ampl})). Also we note that for the chosen parameters we have
$V=0$, so at $x=0$ we have $\kappa=\Omega/\sqrt{gh_0}$ as in linear
theory. It is instructive to compare solutions with ($C_D=0.01$) and
without ($C_D=0$) friction. In Fig.~\ref{fig2} the dependence of the
modulus $m$ on $x$ is shown for both cases.  We see that for the
frictionless case $m\to 1$ with decrease of depth, i.e. the wave
crests assume the shape of solitary waves when one approaches the
shoreline. When $C_D \ne 0$ the modulus also grows with decrease of
depth but never reaches unity. The dependence on $x$ of the mean
surface elevation $\langle A \rangle $ for the cases without and
with friction is shown in Fig.~\ref{fig3}. We have checked that the
``wavelength'' $L$ (\ref{14-2}) is constant for both solutions.
Also, one can see from Fig.~\ref{fig3} (right) that  the value
$h^{1/4}\langle A \rangle \propto d$ is indeed conserved in the
frictionless case but is not constant if  friction is present (the
same holds true for the value $h^{1/2}\langle A^2 \rangle \propto
\sigma$ but we do not present the graph here). Finally, in
Fig.~\ref{fig4} the dependence of the physical elevation wave
amplitude $a$ on the spatial coordinate $x$ is shown. One can see
that the amplitude adiabatically grows with distance in the
frictionless case due to the effect of the slope (without friction)
but, not unexpectedly, gradually decreases in the case when bottom
friction is present, where the decrease for these parameter settings
is comparable in magnitude to the effect of the slope. In both cases
the main qualitative changes occur in the wave shape and the
wavelength.

 Overall, we can infer from these results that the main
local effect of a slope and bottom friction on a cnoidal wave, along
with the adiabatic amplitude variations, is twofold:  a wave with a
$m<1$ at $x=0$ tends to transform into a sequence of solitary waves
as $x$ decreases, and at the same time the distance between
subsequent wave crests tends to decrease. This is in sharp contrast
with the behaviour of modulated cnoidal waves in problems described
by the unperturbed KdV equation, where growth of the modulus $m$ is
accompanied by an {\it increase} of the distance between the wave
crests. Generally, in the study of behaviour of unsteady undular
bores in the presence of a slope and bottom friction we will have to
deal with the combination of these two opposite tendencies.

\section{Undular bore propagation over variable topography with bottom friction}

\subsection{Gurevich-Pitaevskii problem for flat-bottom zero-friction case}

We now turn to the problem (b) outlined in Section 2.  We study the
evolution of an undular bore developing from  an initial surface
elevation jump  $\Delta > 0$,  located at some point $x_0<0$.  As
discussed below, the undular bore will  expand with time so that at
some $t=t_0$ its lead solitary wave enters the gradual slope region,
which begins at $x=0$ (see Fig. 1b). We assume that for $x<0$ one
has $h=h_0=\hbox{constant}$ and $C_D \equiv 0$. We shall first
present a  formulation of the Gurevich-Pitaevskii problem for the
perturbation-free KdV equation and reproduce the well-known
similarity modulation solution describing the evolution of the
undular bore until the moment it enters the slope.  We emphasize
that, although this formulation and, especially, this similarity
solution are known very well and have been used by many authors,
some of the inferences important for the present application to
fluid dynamics have not been widely appreciated, as far as we can
discern. Pertinent to our main objective in this paper, we undertake
a detailed study of the characteristics of the Whitham modulation
system in the vicinity of the leading edge of the undular bore
solution, and show that the boundary conditions of
Gurevich-Pitaevskii type permit only two possible characteristics
configurations, implying two qualitatively different types of the
leading solitary wave behaviour. Next, we shall show how this
Gurevich-Pitaevskii formulation of the problem applies to the
perturbed modulation system in the form (\ref{5-2}) and finally we
will study the effects of the perturbation on the modulations in the
vicinity of the leading edge of the undular bore.

In the case of a flat, frictionless bottom the original equation
(\ref{1-1}) becomes the constant-coefficient KdV equation which can
be cast into the standard form
\begin{equation}\label{kdv0}
\eta_\zeta + 6 \eta \eta_\xi + \eta_{\xi \xi \xi}=0 \,
\end{equation}
by introducing the new variables
\begin{equation}\label{eta}
\eta =\frac{2}{3h_0} A \, , \quad
\xi=\frac{3}{2h_0}(x+x_0-\sqrt{gh_0}t)\, , \quad \zeta
=\frac{9}{16}\sqrt{\frac{g}{h_0}}t \, ,
\end{equation}
where $x_0<0$ is an arbitrary constant. In the Gurevich-Pitaevskii
(GP) approach, one considers a large-scale initial disturbance
$\eta(\xi,0)= f(\xi)$, in the form of  a  decreasing profile,
$f'(\xi) < 0$ (e.g. a smooth step: $f(\xi) \to 0$ as $\xi \to +
\infty$; $f(\xi) \to \eta_0>0$ as $\xi \to -\infty$), whose initial
evolution until some critical (breaking) time $\zeta_b$ can be
described by the dispersionless limit of the KdV equation, i.e. by
the Hopf equation,
\begin{equation}\label{hopf}
\zeta<\zeta_b: \qquad \eta \approx r(\xi, \zeta), \qquad r_\zeta +
6rr_{\xi}=0 \, , \qquad r(\xi,0)=f(\xi)\, .
\end{equation}
The evolution (\ref{hopf}) leads to wave-breaking of the  $r(\xi)$-profile at
some $\zeta=\zeta_b$, with the consequence that  the dispersive term
in the KdV equation then comes into play, and an undular bore forms,
which can be locally represented as a single-phase travelling wave. This
travelling wave is modulated in such a way that it acquires the form
of a solitary wave at the leading edge $\xi = \xi^+(\zeta)$ and gradually
degenerates, via the nonlinear cnoidal-wave regime, to a linear
wave packet at the trailing edge $\xi = \xi^-(\zeta)$. It is important that
this undular bore is essentially unsteady, i.e. the region
$\xi^-(\zeta) < \xi < \xi^+(\zeta)$ expands with time $\zeta$.

The single-phase travelling wave solution of the KdV equation
(\ref{kdv0}) has the form (cf. ~(\ref{cnoidal1}))
\begin{equation}\label{trav}
\eta(\xi, \zeta)= r_3-r_1-r_2 -
2(r_3-r_2)\sn^2(\sqrt{r_3-r_1}\theta, m)
\end{equation}
\begin{equation}\label{}
\theta=\xi+2(r_1+r_2+r_3)\zeta \, , \qquad
m=\frac{r_3-r_2}{r_3-r_1}\, .
\end{equation}
 The parameters $r_1 \le r_2 \le r_3 \le
0$ in the undular bore are slowly varying functions of $\xi, \zeta$,
whose evolution is governed by the Whitham equations
\begin{equation}\label{wh0}
\frac{\partial r_j}{\partial \zeta} +v_j(r_1,r_2,r_3)\frac{\partial
r_j}{\partial \xi}=0 \, , \qquad j=1,2,3.
\end{equation}
The characteristic velocities in ~(\ref{wh0}) are given by
 ~(\ref{5-6}). We stress that, although analytical expressions
(\ref{trav}) and (\ref{cnoidal1}) (as well as ~(\ref{wh0}) and
the homogeneous version of ~(\ref{5-2})) are identical, they are
written for completely different sets of variables, both dependent
and independent.

The Riemann invariants $r_j(\xi, \zeta)$ are subject to special
matching conditions at the free boundaries, $\xi = \xi^{\pm}(\zeta)$
defined by the conditions $m=0$  (trailing edge) and $m=1$ (leading edge),
formulated in Gurevich and  Pitaevskii (1974) (see also
Kamchatnov (2000) or El (2005) for a detailed description).

 At the trailing  (harmonic) edge, where the wave amplitude $a=2(r_3-r_2)$ vanishes and
$m = 0$, one has
\begin{equation}\label{gp1}
\xi=\xi^-(\zeta): \qquad r_2=r_3\, , \quad -r_1=r \, .
\end{equation}
At the leading (soliton) edge, where $m = 1$ one has
\begin{equation}\label{gp2}
\xi=\xi^+(\zeta): \qquad r_2=r_1\, , \quad -r_3=r\, .
\end{equation}
In both ~(\ref{gp1}) and (\ref{gp2}), $r(\xi, \zeta)$ is the
solution of the Hopf equation (\ref{hopf}).

The curves $\xi = \xi^{\pm}(\zeta)$ are defined for the solution of the GP
problem (\ref{wh0}), (\ref{gp1}), (\ref{gp2}) by the ordinary
differential equations
\begin{equation}\label{ode}
\frac{d \xi^-}{d \zeta}= v^-(\xi^-, \zeta)\, , \qquad \frac{d
\xi^+}{d \zeta}= v^+ (\xi^+, \zeta)\, , \qquad
\end{equation}
where $v^{\pm}$ are calculated as the values of  double
characteristic velocities of the modulation system at the undular
bore edges,
\begin{equation}\label{double1}
v^-=v_2(r_1, r_3,r_3) |_{\xi=\xi^-(\zeta)}=  v_3(r_1, r_3,r_3)
|_{\xi=\xi^-(\zeta)},
\end{equation}
\begin{equation}\label{double2}
v^+=v_2(r_1, r_1,r_3) |_{\xi=\xi^+(\zeta)}=v_1(r_1, r_1,r_3)
|_{\xi=\xi^+(\zeta)}
\end{equation}
These equations (\ref{ode}) essentially represent kinematic boundary conditions
for the undular bore (see El 2005). Indeed, the double
characteristic velocity $v_2(r_1, r_3,r_3)=v_3(r_1, r_3,r_3)$ can be
shown to coincide with the linear group velocity of the
small-amplitude KdV wavepacket while the double characteristic
velocity $v_2(r_1, r_1,r_3)=v_1(r_1, r_1,r_3)$ is  the
soliton speed.

One might infer from this GP formulation of the problem that, since
the leading edge of the undular bore specified by ~(\ref{ode}),
(\ref{double2}) is a characteristic of the modulation system, then
the value of the double Riemann invariant $r^+  \equiv r_2=r_1$ is
constant.  Then, on considering an  undular bore propagating into
still water, where $r=0$, one would obtain from the matching
condition (\ref{gp2}) at the leading edge that $r_3|_{\xi=\xi^+}=0$
and thus, the amplitude of the lead solitary wave
$a^+=2(r_3-r_1)|_{\xi = \xi^+}=-r^+$ would always be constant as
well. However, this contradicts the general physical reasoning that
the amplitude of the lead solitary wave should be allowed to change
in the case of general initial data. The apparent contradiction is
resolved by noting that the leading edge specified by ~(\ref{ode}),
(\ref{double2}) can be an {\it envelope} of the characteristic
family, i.e. a caustic, rather than necessarily a regular
characteristic, and hence there is no necessity for the double
Riemann invariant $r^+$ to be constant along the curve $\xi =
\xi^+(\zeta)$ in general case. On the other hand, since the leading
edge is defined by the condition $m=1$, the wave form at the leading
edge will coincide with the spatial profile of the standard KdV
soliton. Thus we arrive at the conclusion that, in general, the
amplitude of the leading KdV solitary wave will vary, even in the
absence of the perturbation terms. Of  course, in the unperturbed
KdV equation, such varying solitary waves  cannot not exist on their
own, and require the presence of the rest of the undular bore. We
also stress that these variations of the leading solitary wave in
the undular bore, as described here, have a completely different
physical nature to the variations of the parameters of an individual
solitary wave due to small perturbations as described in Section 4.
They  are caused by nonlinear wave interactions within the undular
bore rather than by a local adiabatic response of the solitary wave
to a perturbation induced by topography and friction. Importantly
for our study, however, it will transpire that the action of these
same perturbation terms on the undular bore can lead to both a local
and a nonlocal response of the leading solitary wave.

\subsection{Undular bore developing from an initial jump}
Next we consider the simplest solution of the modulation system,
which  describes an undular bore developing from an initial
discontinuity placed at the  point $x=-x_0$. In $(\eta ; \xi,
\zeta)$ - variables we have the initial conditions
\begin{equation}\label{disc}
\eta(\xi, 0)= \Delta \qquad \hbox{for} \quad \xi<0\, ; \qquad
\eta(\xi, 0)=0 \qquad \hbox{for} \quad \xi>0 \, ,
\end{equation}
where $\Delta>0$ is a constant. Then, on using ~(\ref{hopf}), the
initial conditions (\ref{disc}) are readily translated into the
free-boundary matching conditions (\ref{gp1}), (\ref{gp2}) for the
Riemann invariants. Because of the absence of a length
scale in this problem, the corresponding solution of the
modulation system must depend on the self-similar variable $\tau
= \xi/\zeta$ alone, which reduces the modulation system to the
ordinary differential equations
\begin{equation}\label{odegp}
(v_i-\tau)\frac{dr_i}{d\tau}=0\, , \qquad i=1,2,3.
\end{equation}
The boundary conditions for ~(\ref{odegp}) follow from the
matching conditions (\ref{gp1}), (\ref{gp2}) using the
initial condition (\ref{disc}):
\begin{equation}\label{gp}
\begin{split}
&\tau = \tau^-: \qquad r_2 = r_3 \, , \quad r_1 = - \Delta \\
&\tau = \tau^+: \qquad r_2 = r_1 \, , \quad r_3 = 0 \, .
\end{split}
\end{equation}
where $\tau^{\pm}$ are self-similar coordinates (speeds) of the
leading and trailing edges, $\xi^{\pm}=\tau^{\pm}\zeta$. Taking into
account the inequality $r_1\le r_2\le r_3$ one obtains the
well-known modulation solution of Gurevich and Pitaevskii (1974)
(see also Fornberg and Whitham 1978) in the form
\begin{equation}\label{self1}
r_1= - \Delta\, , \quad r_3=0\, , \quad r_2=- m \Delta\,,
\end{equation}
\begin{equation}\label{modsol}
\frac{\xi}{\zeta}=v_2(-\Delta,-m\Delta,0)=
2\Delta[(1+m)-\frac{2m(1-m)K(m)}{E(m)-(1-m)K(m)}]\,.
\end{equation}
\begin{figure}[bt]
\centerline{\includegraphics[width=7cm,height=5cm,clip]{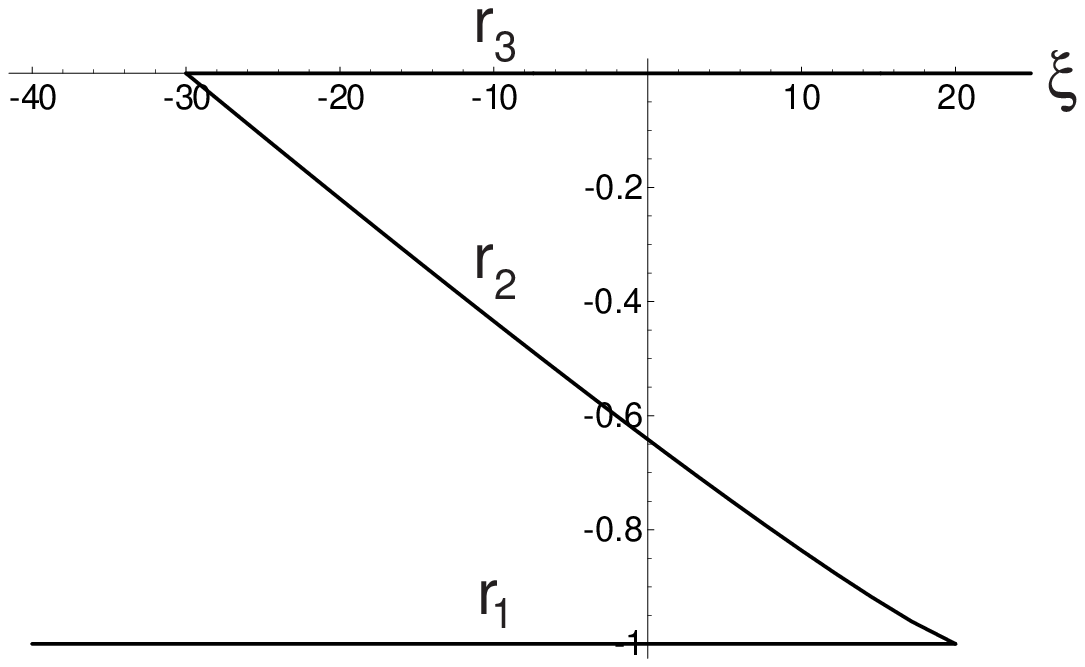}
\includegraphics[width=7cm,height=6cm,clip]{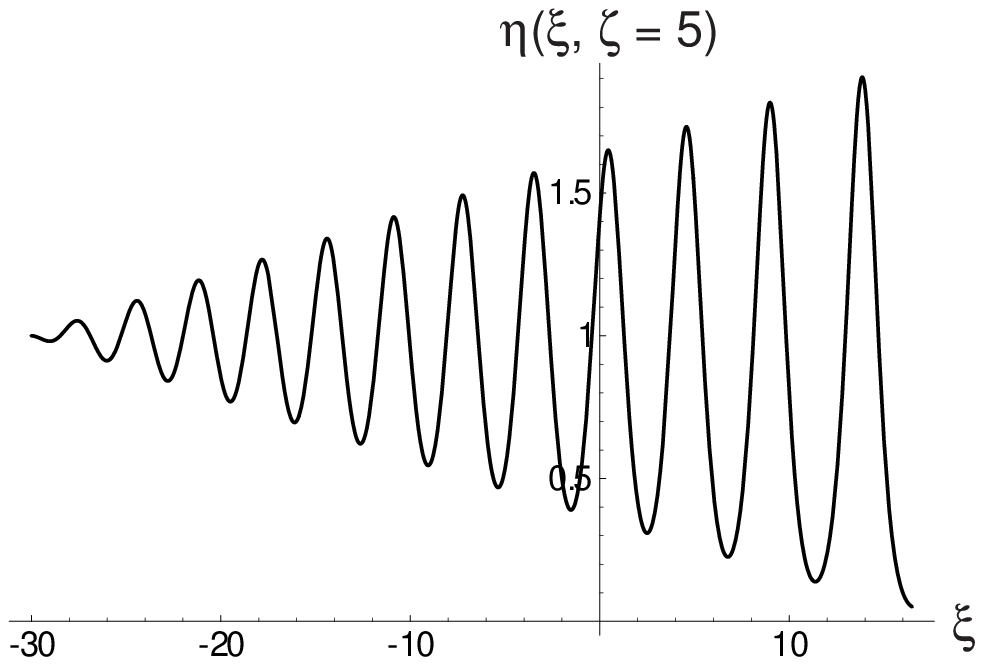}}
\caption{Left: Riemann invariants behaviour in the similarity
modulation solution for the flat-bottom zero-friction case ; \
Right: corresponding undular bore profile $\eta(\xi)$.} \label{fig5}
\end{figure}
This modulation solution (\ref{self1}), (\ref{modsol}) (see Fig.~5a)
represents  the replacement, due to averaging over the oscillations,
of the unphysical formal three-valued solution of the dispersionless
KdV equation (i.e. of the Hopf equation) which would have taken
place in the absence of the dispersive  regularisation by  the
undular bore. We see that ~(\ref{modsol}) describes an expansion fan
in the characteristic $(\xi, \zeta)$-plane and thus is a global
solution. Substituting ~(\ref{self1}), (\ref{modsol}) into the
travelling wave solution (\ref{trav}) one obtains the asymptotic
wave form of the undular bore (see Fig.~5b), which then can be
readily represented in terms of the original physical variables
using the relationships (\ref{eta}).

The equations of the trailing and leading edges of the undular bore
are determined from ~(\ref{modsol}) by putting  $m=0$ and $m=1$
respectively
\begin{equation}\label{edges}
\frac{\xi^- }{\zeta }= \tau^- = v_2(-\Delta,0,0) = -6\Delta\, ,
\qquad \frac{\xi^+ }{\zeta }=\tau^+  = v_2(-\Delta,-\Delta,0)=
4\Delta \, .
\end{equation}
The leading solitary wave amplitude is $\eta_0=2(r_3-r_1)=2\Delta$,
which is exactly twice the height of the initial jump. This
corresponds to the amplitude of the surface elevation $a=3h_0
\Delta$ (see (\ref{eta})).  Note that, to get the leading solitary
wave of the same initial amplitude $a_0$ as for the separate
solitary wave considered in Section 4, one should use the jump value
$\Delta_0=a_0/3h_0 $, which of course is just $2\tilde{\Delta }$,
where $\tilde{\Delta } = 3h_0 \Delta /2$ is the initial discontinuity
in the surface elevation.

\subsection{Structure of the undular bore front}

We  are especially interested in the behaviour of the modulation
solution (\ref{self1}), (\ref{modsol}) in the vicinity of the
leading edge $\xi = \xi^+(\zeta)$.  This behaviour is essentially
determined by the manner in which the pair of characteristics
corresponding to the velocities $v_2$ and $v_1$ merge into a
multiple eigenvalue $v^+$ of the modulation system at
$\xi=\xi^+(\zeta)$.

First, one can readily infer from the modulation solution
(\ref{self1}), (\ref{modsol}) that the phase velocity $c=
-2(r_1+r_2+r_3)=2\Delta(1+m)>v_2(-\Delta, -m, 0)$ for $m<1$ and
$c=v_2$ for $m=1$. Thus, any individual wave crest generated at the
trailing edge of the undular bore moves towards the leading edge,
i.e. for any crest $m \to 1$ as $\zeta \to \infty$. Thus, for any
particular wave crest, except for the very first one, the solitary
wave `status' is achieved only asymptotically as $\zeta \to \infty$.

Without loss of generality we assume in this section that $\Delta=1$
in ~(\ref{self1}), (\ref{modsol}). First, as we have already
mentioned, the characteristic family $\Gamma_2:$ $d\xi/d\zeta=v_2$
is  an expansion fan in the $\xi, \zeta$ - plane,
\begin{equation}\label{}
\Gamma_2: \qquad \xi=C_2 \zeta \,,
\end{equation}
parameterised by a constant $C_2$, $-6 \le C_2 \le 4 $ . Next,  in
 ~(\ref{modsol}) we make an asymptotic expansion of $v_2(-1, -m, 0)$
for small $(1-m) \ll 1$, to get
\begin{equation}\label{84}
2(1-m)\ln ({16}/(1-m)) \simeq\tau^+ - {\xi}/{\zeta}
\end{equation}
or, with logarithmic accuracy,
\begin{equation}\label{mlead}
(\tau^+- {\xi}/{\zeta}) \ll 1 : \qquad
1-m\simeq\frac{\tau^+-\xi/\zeta}{2\ln [1/(\tau^+-\xi/\zeta)]}\, .
\end{equation}
Next, expanding $v_1(- 1, -m, 0)$  for $(1-m)\ll 1$ and using
~(\ref{mlead}) we get the asymptotic equation for the
characteristics family $\Gamma_1$,
\begin{equation}\label{85}
\frac{d \xi}{d \zeta}= v_1  = \tau^+ + (\tau^+ -{\xi}/{\zeta}) +
\mathcal{O}(1-m)\, ,
\end{equation}
which is readily integrated to leading order to give
\begin{equation}\label{g1}
\Gamma_1 : \qquad \xi\simeq \tau^+ \zeta- \frac{C_1}{\zeta} \, ,
\end{equation}
where $C_1\ge 0$ is an arbitrary constant `labeling' the
characteristics; $C_1=0$ corresponds to the leading edge of the
undular bore. This asymptotic formula (\ref{g1}) is valid as long as
$\zeta \gg 1$. The behaviour of the characteristics belonging to the
families $\Gamma_1$ and $\Gamma_2$ near the leading edge is shown in
Fig.~6a.

\begin{figure}[bt]
\centerline{
\includegraphics[width=5cm,height=5.5cm,clip]{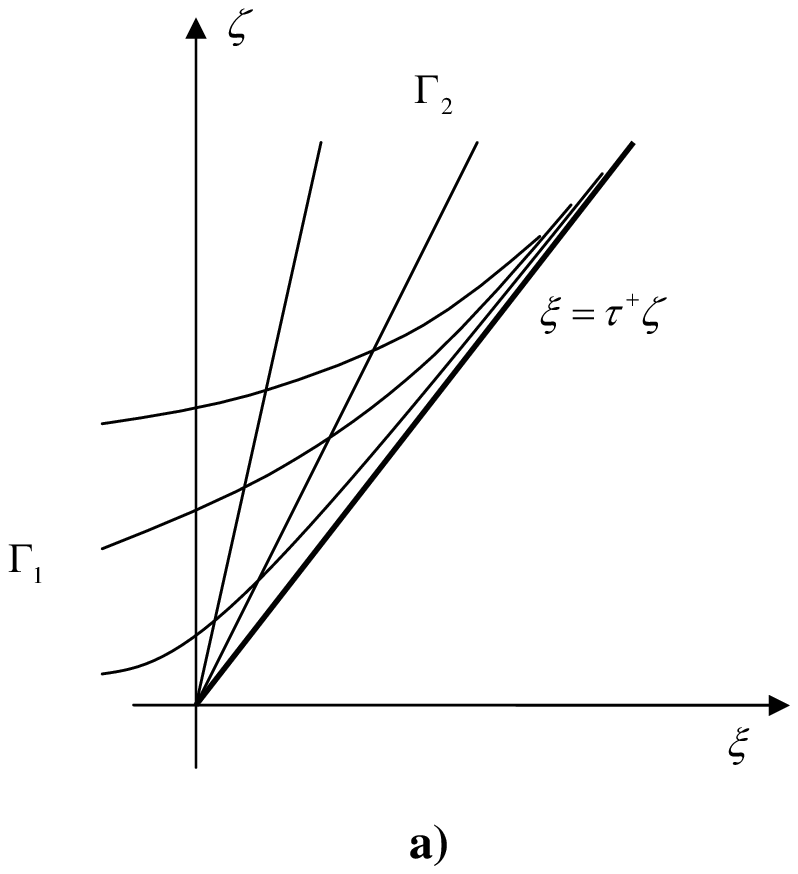} \qquad
\qquad
\includegraphics[width=4.5cm,height=5.5cm,clip]{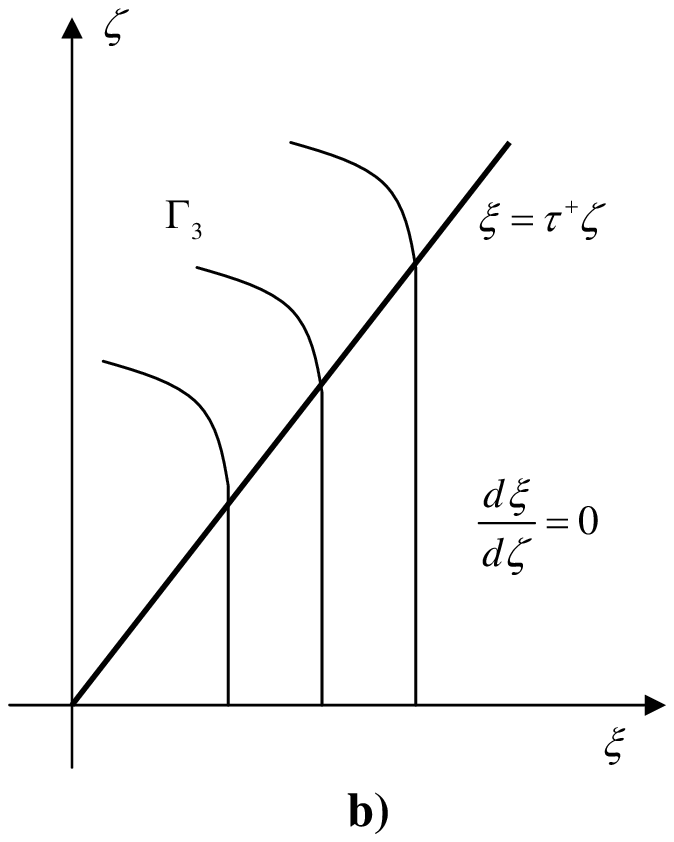}
} \caption{Characteristics behaviour for the similarity modulation
solution near the leading edge $\xi^+(\zeta)$: (a) families
$\Gamma_1$: $ d\xi/d\zeta = v_1 $ and $\Gamma_2$ : $\xi=C_2 \zeta$,
\ (b) family $\Gamma_3$: \ $d\xi/d\zeta = v_3$.  } \label{fig6}
\end{figure}

Next, expanding the equation for the third characteristic family,
$\Gamma_3$: $d\xi/ d \zeta=v_3(-1, -m, 0)$ for $(1-m) \ll 1$, we get
on using (\ref{mlead})
\begin{equation}\label{g3}
\frac{d \xi}{d \zeta} = \frac{\tau^+-\xi/\zeta}{\ln(1/(\tau^+ -
\xi/\zeta))} + \mathcal{O}(\tau^+ - \xi/\zeta)\, .
\end{equation}
Integrating (\ref{g3}) we obtain to first order
\begin{equation}\label{88}
\Gamma_3: \qquad \xi \simeq C_3 - g(\zeta) \, ,
\end{equation}
\begin{equation}\label{}
\hbox{where} \quad g(\zeta)= \int \frac{1}{\zeta}\frac{\tau^+ \zeta
-C_3}{\ln |\tau^+ \zeta - C_3|-\ln \zeta} d\zeta\, , \qquad
g(C_3/\tau^+)=0 \, ,
\end{equation}
$C_3$ being an arbitrary constant. The asymptotic formula (\ref{88})
is valid as long as $g(\zeta)/C_3 \ll 1$. Since the characteristics
$\Gamma_3$ intersect the leading edge $\xi = \tau^+ \zeta$ we must
indicate their behaviour outside the undular bore. It follows from
the matching condition (\ref{gp2}) and the limiting structure
(\ref{6-2}) of the characteristic velocities of the Whitham system,
that the characteristics from the family $\Gamma_3$ match with the
Hopf equation characteristics $d\xi/d\zeta=6r$ carrying the value of
the Riemann invariant $r=0$ corresponding to still water upstream
the undular bore. Therefore, the sought external characteristics are
simply vertical lines $\xi=C_3$.  The qualitative behaviour of the
characteristics from the family $\Gamma_3$ is shown in Fig.~6b.

It is clear from the asymptotic behaviour of the
characteristics that the edge characteristic $\xi= \tau^+ \zeta$
corresponding to the motion of the leading solitary wave intersects
only with characteristics of the family $\Gamma_3$ carrying the
Riemann invariant value $r_3=0$ into the undular bore domain.
Since, according to the matching condition (\ref{gp}), $r_3 \equiv 0$
everywhere along the edge characteristic one can infer that the leading
solitary wave motion is completely specified by its amplitude at
$\zeta=0$. Indeed, in this case, the leading edge represents a genuine multiple
characteristic of the modulation system, along which the Riemann
invariant $r^+=r_2=r_1$ is a constant. Given the constant value of
$r_1=-1$ for the solution (\ref{modsol}), one infers that the
amplitude of the lead soliton of the self-similar undular bore,
$\eta_0=2(r_3-r^+)=2$ is also a constant value. Thus, in the undular
bore evolving from an initial jump, the leading solitary wave
represents an independent soliton of the KdV equation. Of course,
this fact follows directly from the modulation solution
(\ref{modsol}) but now we have established its meaning in the
context of the characteristics, which will play an important
role below.

Next  we  discuss the structure of the undular bore front in the
case when the initial profile $\eta(\xi, 0)$ is not a simple  jump
discontinuity, and instead  has the form of a monotonically
decreasing function, for instance, $(-\xi)^{1/2}$ when $\xi \le0$
and $\eta(\xi, 0)=0$ for $\xi
>0$. In that case, the modulation solution for the undular bore no
longer possesses  $x/t$-similarity as in the jump resolution case
and, as a result, the speed (and therefore, the amplitude) of the
lead solitary wave is not constant.  For instance, for the
afore-mentioned square-root initial profile the amplitude of the
lead solitary wave grows as $\zeta^{2}$ (see Gurevich, Krylov and
Mazur 1989, or Kamchatnov 2000). Clearly, such an amplitude
variation is
 impossible if the leading edge $\xi^+(\zeta)$ was a
regular characteristic carrying a constant value of the Riemann
invariant $r^+$.  As discussed above, however, the
GP matching conditions (\ref{gp1}) -(\ref{double2})
admit another possibility; the leading  edge curve is the
 {\it envelope} of the characteristic families
$\Gamma_1$: $d\xi/d\zeta=v_1$ and $\Gamma_2$: $d\xi/d\zeta=v_2$
merging when $m=1$. This configuration is shown in Fig. 7a. In this
case, the behaviour of the modulus $m$ in the vicinity of the
leading edge is given by the asymptotic formula found in  Gurevich
$\&$ Pitaevskii (1974):
\begin{figure}[bt]
\centerline{
\includegraphics[width=5cm,height=5.5cm,clip]{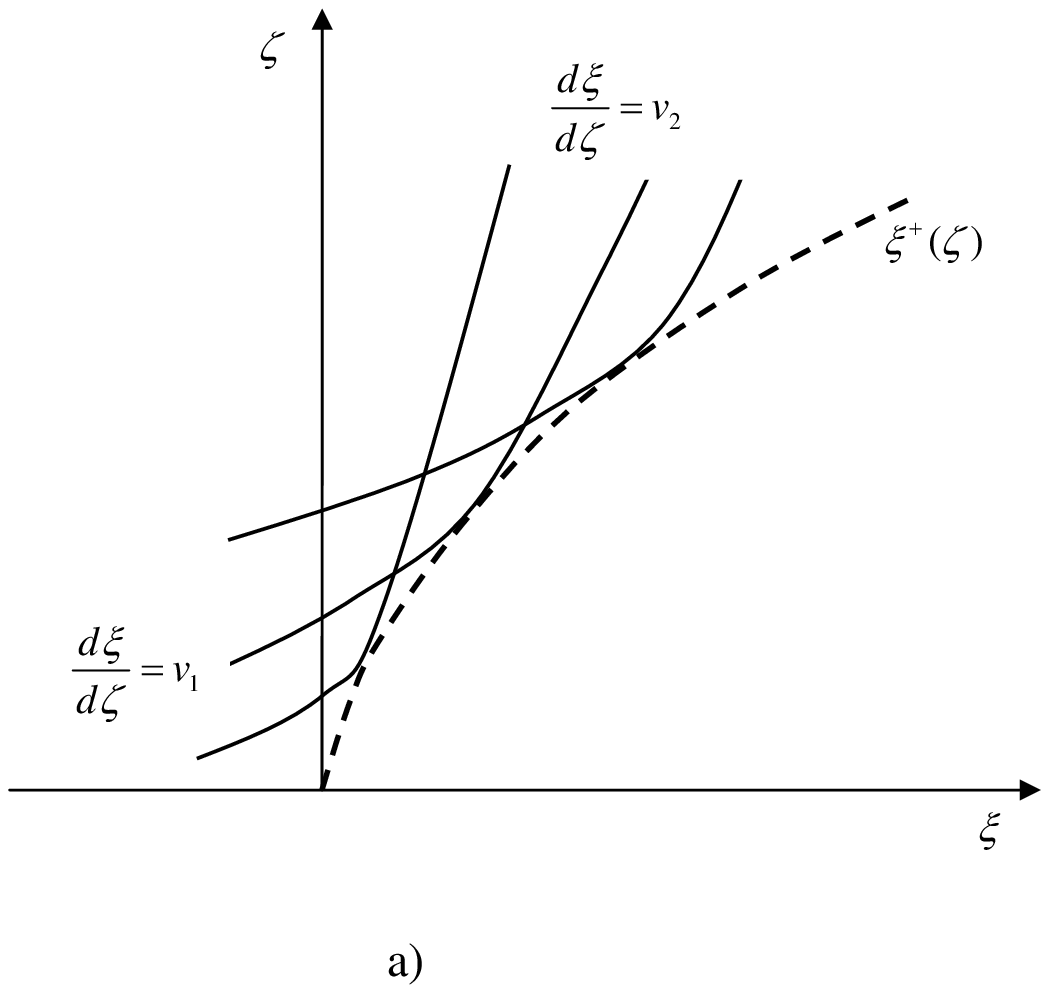} \qquad
\qquad
\includegraphics[width=4.5cm,height=3.5cm,clip]{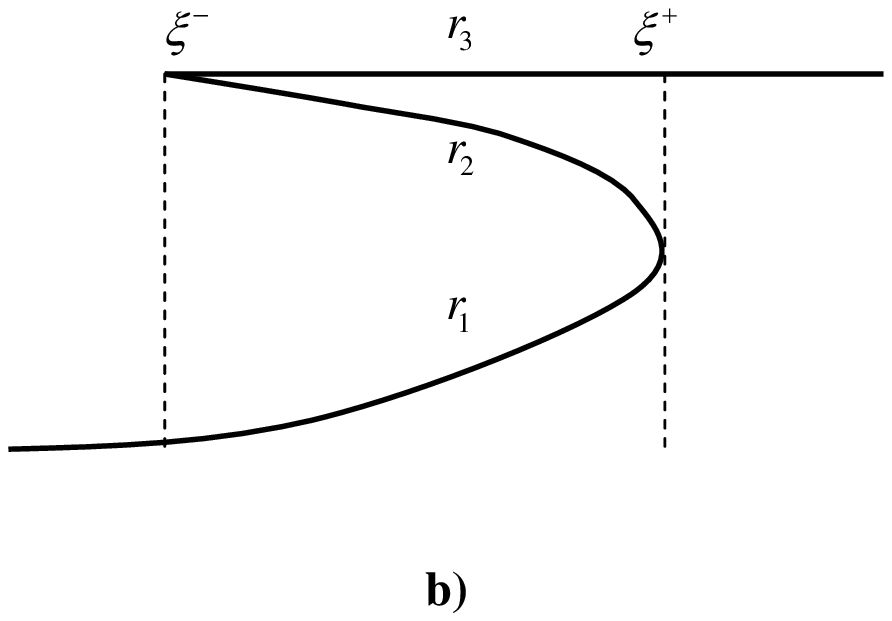}
} \caption{a) Leading edge $\xi^+(\zeta)$ of non-self-similar
undular bore as an envelope of pairwise merging characteristics from
the families $d\xi/d\zeta=v_1$ and $d\xi/d\zeta=v_2$; \ b) behaviour
of the Riemann invariants in non-self-similar modulation solution
with $r_3 \equiv 0$. } \label{fig7}
\end{figure}
\begin{equation}\label{82}
(1-m)^2 \left( \ln \frac{16}{1-m} + \frac{1}{2}
\right)=\frac{2}{(r^+)^2}\frac{d r^+}{d \zeta}(\xi^+-\xi)
\end{equation}
where the function $r^+(\zeta) \ne \hbox{constant}$ is assumed to be
known. Another specific feature of this (general) configuration is
that $dr_{1,2}/d\xi \to \pm \infty$ as $\xi \to \xi^+$ (see Fig. 7b
- also found in Gurevich $\&$ Pitaevskii 1974, see also Kamchatnov
2000), which is in drastic contrast with similarity solution (see
Fig. 6a). This particular difference was discussed in relation with
undular bores in the KdV-Burgers equation in Gurevich and Pitaevskii
(1987).

In summary, we see from ~(\ref{82}) that the structure of the
modulation solution in the vicinity of the leading edge of an
undular bore defined as a characteristic envelope is qualitatively
different compared to that for the similarity case (see
~(\ref{84})). The more general (but qualitatively similar to
(\ref{82})) asymptotic formula which takes into account small
perturbations due to a variable topography and bottom friction will
be derived later. At the moment, it is important for us that in this
configuration, when the leading edge is a characteristic envelope
rather than just a characteristic, the value $r^+$, and thus, the
leading solitary wave amplitude  are allowed to vary.

The analysis of the corresponding modulation solution in Gurevich,
Krylov and Mazur (1989) showed that, while in the case of an initial
jump the wave crests generated at the trailing edge reach the
leading edge (and therefore, transform into solitary waves) only
asymptotically as $t \to \infty$, for the more general case of
decreasing initial data each wave crest generated at the trailing
edge reaches the leading edge in finite time and replaces
(overtakes) the existing leading solitary wave. This process is
manifested as a continuous amplitude growth of the (apparent)
leading solitary wave. As in classical soliton theory, an
alternative explanation of the leading solitary wave amplitude
growth can be made in terms of the momentum exchange between the
``instantaneous'' leading solitary wave and solitary waves of
greater amplitude coming from the left. Indeed, as the rigorous
analysis of Lax, Levermore and Venakides showed (see Lax, Levermore
and Venakides (1994) and the references therein), the whole
modulated structure of the undular bore can be asymptotically
described in terms of the interactions of a large number of KdV
solitons initially `packed' into a non-oscillating large-scale
initial profile.

This latter interpretation is especially instructive for our
purposes.  Our point is that the specific cause of the enhanced
soliton interactions resulting in  amplitude growth at the leading
edge is not essential;  it can be large-scale spatial variations of
the initial profile as  just described, but it could also equally
well be an effect of  small perturbations in the KdV equation
itself. Indeed, in the weakly perturbed KdV equation, the local wave
structure of the undular bore must be described to leading order by
the periodic solution (\ref{trav}) of the {\it unperturbed} KdV
equation, so if one assumes the GP boundary conditions analogous to
(\ref{gp1}) -- (\ref{double2}) for the perturbed modulation system
(\ref{5-2}), one invariably will have to deal with one of the two
possible types of the characteristics behaviour (shown in Figs. 7a
and 8a) in the vicinity of the leading edge of the undular bore,
because this qualitative behaviour is determined only by the
structure of the GP boundary conditions and by the associated
asymptotic structure of the characteristic velocities of the Whitham
system for $(1-m)\ll 1$, which are the same for both unperturbed and
perturbed modulation systems. Next, we will show that, by using the
knowledge of this qualitative behaviour of the characteristics, one
is able to construct the asymptotic modulation solution for the
undular bore front in the presence of variable topography and bottom
friction even if the full solution of the perturbed modulation
system is not available.

\subsection{Gurevich-Pitaevskii problem for perturbed modulation system}

We investigate now how the GP matching problem  applies to the
perturbed modulation system (\ref{5-2}). As in the original GP
problem, we postulate the natural physical requirement that the mean
value $\langle U \rangle$ is continuous across the undular bore
edges, which represent  free boundaries and are defined by the
conditions $m=0$ (trailing edge $X=X^{-} (T)$) and $m=1$ (leading
edge $X=X^{+}(T)$). Also, we consider propagation of the undular
bore into still water, hence $\langle U \rangle |_{X=X^{+}(T)}=0$.
Now, using the explicit expression (\ref{avU}) for $\langle U
\rangle$ in terms of complete elliptic integrals and calculating its
limits as $m \to 0$ and $m \to 1$ one has
\begin{equation}\label{GP1}
\begin{split}
X= X^-(T): \qquad \la_2=\la_3\, ,\quad  \langle U \rangle = -\la_1=u \, ,  \\
X= X^+(T): \qquad \la_2=\la_1\, , \quad \langle U \rangle =  -\la_3=0 \, ,  \\
\end{split}
\end{equation}
where $u(X,T)$ is solution of the dispersionless perturbed KdV
equation (\ref{U}), i.e.
\begin{equation}\label{6-11}
    u_T+6uu_X=F(T)u-G(T)u^2,
\end{equation}
with the boundary conditions
\begin{equation}\label{bcU}
u\left (\tau, \frac{1}{6g}\int^{\tau}_0h d\tau \right )=
\frac{9g}{2h_0}\Delta_0 \quad \hbox{if}\quad \tau< \tau_0; \qquad u
\left(\tau, \frac{1}{6g}\int^{\tau}_0hd\tau \right )= 0 \quad
\hbox{if}\quad \tau> \tau_0 \, ,
\end{equation}
where $\tau_0=-x_0/\sqrt{gh_0}$. The boundary conditions (\ref{bcU})
correspond to a discontinuous initial surface elevation $A(x,t)$ at $x=-x_0$,
  obtained by using transformations  (\ref{1-3}) and (\ref{newvar})
  where one sets $t=0$. As earlier,
$\Delta_0=a_0/(3h_0)$  is the value of the discontinuity in $A$,
chosen in such a way that the amplitude of the lead solitary wave in
the undular bore was exactly $a_0$ in the flat-bottom zero-friction
region (see Section 6.2).

This  free-boundary matching problem is then
complemented by the kinematic conditions explicitly defining the
boundaries $X=X^{\pm}(T)$. These are formulated using the multiple
characteristic directions of the perturbed modulation system
(\ref{5-2}) in the limits as $m \to 0$ and $m \to 1$  (cf. ~
(\ref{ode}) - (\ref{double2})),
\begin{equation}\label{ode1}
\frac{d X^-}{d T}= V^-(X^-, T)\, , \qquad \frac{ d X^+}{d T}= V^+
(X^+, T)\, , \qquad
\end{equation}
\begin{equation}\label{double11}
\hbox{where}  \quad V^-=v_2(u, \la^-,\la^-) =  v_3(u, \la^-, \la^-) ,
\end{equation}
\begin{equation}\label{double21}
V^+=v_2(\la^+, \la^+, 0) =v_1(\la^+, \la^+, 0) \, ,
\end{equation}
\begin{equation}\label{}
\hbox{and} \quad \la^-=\la_2(X^-, T)=\la_3(X^-, T)\, , \qquad \la^+=\la_2(X^+,
T)=\la_1(X^+,T).
\end{equation}
Thus, for the perturbed KdV equation  the leading and trailing edges
of the undular bore are defined mathematically in the same way as
for the unperturbed one, albeit for a different set of variables.

\subsection{Deformation of the undular bore front due to variable topography and bottom friction}

Finally we study the effects of gradual slope and bottom friction on
the leading front of the self-similar expanding undular bore
described in Sections 6.2, 6.3. The result will essentially depend
on the relative values of the small parameters appearing in the
problem. We note that in general  there are three distinct relevant
small parameters,
\begin{equation}\label{}
\epsilon = \frac{h_0}{x_0} \ll 1 \, , \qquad \delta = \max (h_x) \ll
1, \qquad C_D \ll 1
\end{equation}
The first small parameter is determined by the ratio of the
equilibrium depth in the flat bottom region, to the distance from
the beginning of the slope region to the location of the initial
jump discontinuity in the surface displacement. This measures the
typical relative spatial variations of the modulation parameters in
the undular bore when it reaches the beginning of the slope. The
second and third parameters are contained in the KdV  equation
(\ref{1-1}) itself and measure the values of the slope and bottom
friction respectively. In terms of the transformed variables
appearing in ~(\ref{U}), $ |F(T)| \sim \delta$, $|G(T)| \sim C_D$
(see ~(\ref{F})). Generally we assume $\delta \sim C_D$ (the
possible orderings $\delta \ll C_D$ or $C_D \ll \delta$ can be then
considered as particular cases).

To obtain a  quantitative description  of the vicinity of the
leading edge of the undular bore we perform an expansion of the
Whitham modulation system (\ref{5-2}) for $(1-m) \ll 1$. We first
introduce the substitutions
\begin{equation}\label{li}
\la_i(X,T)=\la^+(T)+l_i(\tilde X, T)\, , \quad v_i = V^+ +v'_i \, ,
\quad  \rho_i=\rho^+ + \rho'_i,  \quad i=1,2.
\end{equation}
\begin{equation}\label{v+}
\hbox{where} \quad \tilde X = X^+-X\, , \qquad  V^+=-4\la^+ \, , \quad
\rho^+=\frac43F(T)\la^++\frac{32}{15}G(T)(\la^+)^2.
\end{equation}
Since $\la_2 \ge \la_1$, $v_2 \ge v_1$ one always has $l_2 \ge l_1$,
$v_2' \ge v_1'$. Assuming $\tilde X/X^+ \ll 1 \Leftrightarrow 1-m
\ll 1$ and using that $\la_3=0$ to leading order in the vicinity of
the leading edge (see the matching condition (\ref{GP1})), we have
from asymptotic expansions of ~(\ref{5-3}) -- (\ref{5-6}) as $(1-m)
\ll 1$
\begin{equation}\label{11-1}
    \begin{split}
    v_1'&=M_1(l_2-l_1)\equiv -2\left[1+\frac{\ln({16}/{(1-m)})}{1+\frac{1}{4}(1-m)\ln({16}/{(1-m)})} \right](l_2-l_1),\\
    v_2'&=M_2(l_2-l_1)\equiv -2\left[1-\frac{\ln({16}/{(1-m)})}{1-\frac{1}{4}(1-m)\ln({16}/{(1-m)})} \right](l_2-l_1),
    \end{split}
\end{equation}
\begin{equation}\label{11-2}
    \begin{split}
    \rho_1'=N_1(l_2-l_1)&\equiv\Bigg\{-\frac13\left[1+\ln\frac{l_2-l_1}{-16 \la^+}\right]F\\
    &-\frac4{15}\left[2\la^+\ln\frac{l_2-l_1}{-16\la^+}
    -3\la^+\right]G\Bigg\}(l_2-l_1)\\
    \rho_2'=N_2(l_2-l_1)&\equiv\Bigg\{\frac13\left[5+\ln\frac{l_2-l_1}{-16\la^+}\right]F\\
    &+\frac4{15}\left[2\la^+\ln\frac{l_2-l_1}{-16\la^+}
    +13\la^+\right]G\Bigg\}(l_2-l_1).
    \end{split}
\end{equation}
Naturally, $v_i'$ and $\rho_i'$ vanish when $l_2=l_1$. Now,
substituting ~(\ref{li}), (\ref{v+}) into the modulation system
(\ref{5-2}) we obtain
\begin{equation}\label{10-1}
    \frac{d\la^+}{dT} + \frac{\prt l_i}{\prt \tilde X}\frac{dX^+}{dT}-(V^++v_i')\frac{\prt l_i}{\prt \tilde X} =
    \rho^++\rho'_i, \quad i=1,2.
\end{equation}

On using the kinematic condition (\ref{ode1}) at the leading edge,
this reduces to
\begin{equation}\label{10-1a}
    \frac{d\la^+}{dT}-v_i'\frac{\prt l_i}{\prt \tilde X} =
    \rho^++\rho'_i, \quad i=1,2.
\end{equation}
There are two qualitatively different cases to consider:

(i) $\lim_{\tilde X \to 0}|dl_i/d\tilde X|< \infty $, \ \ $i=1,2$
(Fig. 8a)

(ii) $\lim_{\tilde X \to 0}|dl_i/d\tilde X|= \infty $, \ \ $i=1,2$
(Fig. 8b)
\begin{figure}[bt]
\centerline{
\includegraphics[width=5cm,height=5.5cm,clip]{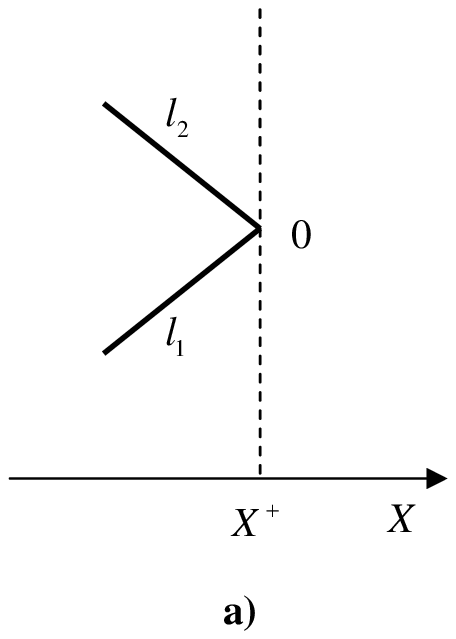} \qquad
\qquad
\includegraphics[width=4.5cm,height=5.5cm,clip]{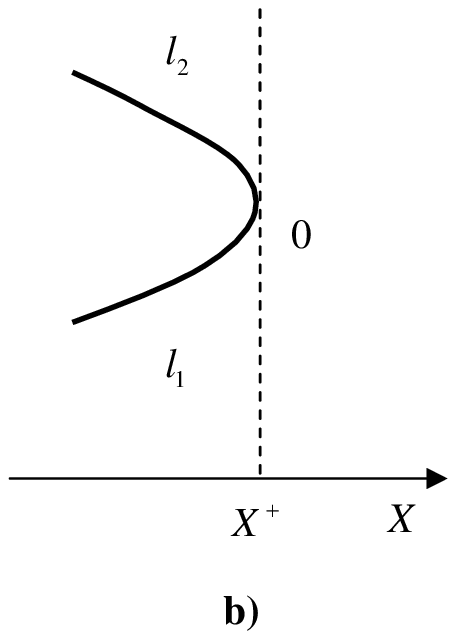}
} \caption{Riemann variables behaviour in the vicinity of the
leading edge of the undular bore propagating over gradual slope with
bottom friction (a) Adiabatic variations of the similarity  GP
regime, $\delta \ll \epsilon$, $C_D \ll \epsilon$; \   (b) General
case, $\delta \sim C_D \sim \epsilon$.} \label{fig8}
\end{figure}

\noindent The case (i) implies that to leading order
~(\ref{10-1a}) reduces to
\begin{equation}\label{lord}
\frac{d\la^+}{dT}=\rho^+ \, ,
\end{equation}
which, together with the kinematic condition $dX^+/dT=-4\la^+$,
defines the leading edge curve $X^+(T)$. One can observe that this
system coincides with ~(\ref{8-3}), (\ref{8-2}) defining the
motion of a separate solitary wave over a gradual slope with bottom
friction. Its integral expressed in terms of original physical
$x,t$-variables is given by ~(\ref{8-8}). Therefore, in the case
(i) the lead solitary wave in the undular bore to leading order is
not restrained by interactions with the remaining part of the bore
and behaves as a separate solitary wave. Physically this case
corresponds to adiabatic deformation of the similarity modulation
solution (\ref{self1}), (\ref{modsol})  and implies the following
small parameter ordering : $\delta \ll \epsilon$, $C_D \ll
\epsilon$.

Next, we study the structure of this weakly perturbed similarity
modulation solution in the vicinity of the leading edge. The next
leading order of the system (\ref{10-1a}) yields
\begin{equation}\label{}
-v_i'\frac{\prt l_i}{\prt \tilde X} = \rho'_i, \quad i=1,2,
\end{equation}
that is
\begin{equation}\label{11-4}
    \frac{\prt l_1}{\prt \tilde X}=-\frac{N_1}{M_1},\qquad
    \frac{\prt l_2}{\prt \tilde X}=-\frac{N_2}{M_2}.
\end{equation}
Subtraction of one equation (\ref{11-4}) from another with account
of the relationship $l_2-l_1\cong-\la^+(1-m)$  leads consistently to
leading order to the differential equation for $1-m$
\begin{equation}\label{11-5}
    \frac{\prt(1-m)}{\prt \tilde X}=2 \left[\frac{F(T)}{-3\la^+}-
    \frac{16 G(T)}{15}\right]
    \left(\ln\frac{16}{1-m}\right)^{-1},
\end{equation}
This equation should be solved with the initial condition
\begin{equation}\label{11-6}
    1-m=0\quad \mathrm{at} \quad \tilde X=0 \, .
\end{equation}
Elementary integration gives with the accuracy $\mathcal{O}(1-m)$
(cf. (\ref{84}))
\begin{equation}\label{12-1}
    (1-m)\ln{\frac{16}{1-m}}=-2\left[\frac13F(T)-\frac{16}{15}\la^+
    G(T)\right]\frac{X^+-X}{-\la^+}.
\end{equation}
This formula determines the dependence of the modulus $m$  on $T$
and $X$ (as long as $1-m\ll1$).

Now, we make use of the solution  $\la^+$ of equation (\ref{lord})
given by ~(\ref{8-6}) with $C_0=4/(3ga_0h_0)$ (see ~(\ref{8-7})).
Under supposition that the integral $\int^xh^{-3}dx$ diverges as
$h\to0$, so that the turbulent bottom friction plays an essential
role in the undular bore front behaviour (see Section 4 for a
similar approximation for an isolated solitary wave), we obtain for
$h \ll h_0$
\begin{equation}\label{12-2}
(1-m)\ln\frac{16}{1-m}=\frac{64}{15}C_D\left(2+3h^2\int_0^x\frac{dx}{h^3}
\right)(X^+-X).
\end{equation}
At last, if the bottom topography is approximated by the dependence
(\ref{9-1}), we get with the same accuracy
\begin{equation}\label{12-3}
    (1-m)\ln\frac{16}{1-m}=\frac{64}{15}C_D\left[2+\frac3{(3\al-1)\delta}
    \left(\frac{h}{h_0}\right)^{1/\al}\right](X^+-X) \, ,
\end{equation}
where $\alpha > 1/3$.  The second term in square brackets tends to
zero as $h\to0$. However, the region where it can be neglected may
be very narrow because of smallness of the parameter $\delta$. We
recall that in this formula $X^+$ is given by ~(\ref{8-8}) and $X$
is defined by ~(\ref{1-3}) in terms of the original physical
independent variables $x$ and $t$.

Summarising, if the conditions $\delta, C_D \ll \epsilon$ are
satisfied, the lead solitary wave of the undular bore behaves as an
individual (noninteracting) solitary wave adiabatically varying
under small perturbation due to variable topography and bottom
friction. The modulation solution in the vicinity of the leading
edge also varies adiabatically, however, its {\it qualitative}
structure considered in Section 6.4 (see Figs~5,6) remains
unchanged.

In a sharp contrast with the  described case of adiabatic
deformation of an undular bore front is case (ii) when the second
term in the left-hand side of ~(\ref{10-1a}) contributes to the
leading order, i.e. to the motion of the leading edge itself.
Namely, we have
\begin{equation}\label{10-1b}
\frac{d\la^+}{dT} =
    \rho^+ + v_i'\frac{\prt l_i}{\prt \tilde X}, \quad i=1,2.
\end{equation}
Now $d\la^+/dT \ne \rho^+$ which means that the amplitude of the
lead solitary wave $a=-2\la^+$ varies essentially differently
compared to the case of an isolated solitary wave. Indeed, the term
$\rho^+$ in the right-hand side of ~(\ref{10-1b}) is responsible for
local adiabatic variations of the solitary wave while the term
$v_i'{\prt l_i}/{\prt \tilde X}$ describes nonlocal parts of the
variations associated with the wave interactions within the undular
bore. Using asymptotic formulae (\ref{11-1}) implying $v'_2 \ge 0$,
$v'_1 \le 0$, and the condition $\lim_{\tilde X \to
0}|dl_{1,2}/d\tilde X|= \infty $ along with $l_2 \ge l_1$, it is not
difficult to show that this nonlocal term is always nonnegative ,
i.e. the lead solitary wave in the undular bore propagating over a
gradual slope with bottom friction always moves faster (and,
therefore, has greater amplitude) than an isolated solitary wave of
the same initial amplitude in the beginning of the slope.  Indeed,
as we have shown in Section 5, the presence of a slope and bottom
friction always result in ``squeezing'' the cnoidal wave, hence
increasing momentum exchange between solitary waves in the vicinity
of the leading edge of the undular bore and acceleration of the lead
solitary wave itself. The situation here is qualitatively analogous
to that described in Section 6.4 where the general global modulation
solution for the unperturbed KdV equation was discussed. Similarly
to that case, the leading edge now represents a characteristic
envelope -- a caustic (otherwise we are back in the case (i)
implying $d\la^+/dT=\rho^+$) (see Fig.~6a).

Unlike the case of adiabatic variations of the leading edge,
determination of the function $\la^+(T)$ requires now knowledge of
the full solution of the perturbed modulation system (\ref{5-2})
with the matching conditions (\ref{GP1}). While the analytic methods
to construct such a solution for inhomogeneous quasilinear systems
are not available presently, it is instructive to assume that
$d\la^+/dT-\rho^+$ is a known function of $T$ and to study the
structure of the solution in close vicinity of the leading edge.
With an account of the
 explicit form (\ref{11-1}) of the velocity corrections, equations (\ref{10-1b}) assume the form
\begin{equation}\label{13-1}
   \frac{\prt l_2}{\prt  \tilde X} =-\frac{d\la^+/dT-\rho^+}
    {2(l_2- l_1)}\left[\frac1{\ln[16/(1-m)]}+\frac14(1-m)\right],
\end{equation}
\begin{equation}\label{13-11}
   \frac{\prt l_1}{\prt \tilde X}=-\frac{d\la^+/dT-\rho^+}
    {2(l_2-l_1)}\left[-\frac1{\ln[16/(1-m)]}+\frac14(1-m)\right].
\end{equation}
Taking the difference of ~(\ref{13-1}) and (\ref{13-11}) we
transform it to the form
\begin{equation}\label{13-2}
    \frac{\prt(1-m)}{\prt X}=\frac{d\la^+/dT-\rho^+}{(\la^+)^2}
    \cdot\frac1{(1-m)\ln[16/(1-m)]}.
\end{equation}
This equation can be readily  integrated with the initial condition
(\ref{11-6}) to give
\begin{equation}\label{13-3}
    (1-m)^2\left(\ln{\frac{16}{1-m}}+\frac12\right)=\frac{2(d\la^+/dT-\rho^+)}
    {(\la^+)^2}(X^+-X).
\end{equation}
This solution coincides  with the asymptotic formula (\ref{82}) for
the behaviour of the modulus in the vicinity of the leading edge of
the undular bore in general unperturbed GP  problem \cite{GP1} but
instead of the derivative $d\lambda^+/dT$ in (\ref{82}) we have the
difference $d\lambda^+/dT - \rho^+$ (which is always positive as we
have established).

\section{Conclusions}

We have studied the effects of a gradual slope and turbulent
(Chezy) bottom friction on the propagation of solitary waves,
nonlinear periodic waves and undular bores in shallow-water flows
in the framework of the variable-coefficient perturbed KdV equation. The analysis has
been performed in the most general setting provided by the
associated Whitham equations describing slow modulations of a
periodic travelling wave due to the slope, bottom friction and
spatial nonuniformity of initial data. This modulation theory,
developed in general form for perturbed integrable equations in
Kamchatnov (2004) was applied here to the perturbed KdV equation and
allowed us to take into account slow variations of all three
parameters in the cnoidal wave solution. The particular
time-independent solutions of the perturbed modulation equations
were shown to be consistent with the adiabatically varying solutions
for a single solitary wave and for a periodic wave propagating over
a slope without bottom friction obtained in Ostrovsky $\&$
Pelinovsky (1970, 1975) and Miles (1979, 1983a). It was shown,
however, that the assumption of zero mean elevation used in these
papers for the description of slow variations of a cnoidal wave,
ceases to be valid in the case when the turbulent bottom friction is
present. In this case, a more general solution was obtained
numerically improving the results of Miles (1983b).

Further, the derived full time-dependent modulation system was used
for the description of the effects of variable topography and bottom
friction on the propagation of undular bores, in particular on the
variations of the undular bore front representing a system of weakly
interacting solitary waves. By the analysis of the characteristics
of the Whitham system in the vicinity of the leading edge of the
undular bore, two possible configurations have been identified
depending on whether the leading edge of the undular bore represents
a regular characteristic of the modulation system or its singular
characteristic, i.e. a caustic. The first case was shown to
correspond to adiabatically slow deformations of the classical
Gurevich-Pitaevskii modulation solution and is realised when the
perturbations due to variable topography and bottom friction are
small compared with the existing spatial non-uniformity of
modulations in the undular bore (which is supposed to be formed
outside the region of variable topography/bottom friction). In the
case when modulations due to the external perturbations  are
comparable in magnitude with the existing modulations in the undular
bore, the leading edge becomes a caustic, and this situation was
shown to correspond to enhanced solitary wave interactions within
the undular bore front. These enhanced interactions have been shown
to lead to a ``nonlocal'' leading solitary wave amplitude growth,
which cannot be predicted in the frame of the traditional local
adiabatic approach to propagation of an isolated solitary wave in a
variable environment.  As we mentioned in the Introduction, one of
our original motivations for this study was the possibility to model
a shoreward propagating tsunami as an undular bore.  In this
context, we would suggest that the second scenario described above
is the more relevant, which has the implication that the growth, and
eventual breaking of the leading waves in a tsunami wavetrain,
cannot be modeled as a local effect for that particular wave, but is
determined instead by the whole structure of the wavetrain.

\section*{Acknowledgements}

This work was started during the visit of A.M.K. at the Department
of Mathematical Sciences, Loughborough University, UK. A.M.K. is
grateful to EPSRC for financial support.

\bigskip

\appendix{\bf Appendix A: Derivation of the perturbed modulation system}

We express the integrand function in the right-hand side of
(\ref{pertmod}) in terms of the $\mu$-variable (\ref{2-5}):
\begin{equation}\label{3-4}
    \begin{split}
    &(2\la_i-s_1-U)R= 8G\mu^3-[8G\la_i+4(F+2s_1G)]\mu^2\\
    &+[4(F+2s_1G)\la_i+2s_1(s_1G+F)]\mu
    -2s_1(s_1G+F)\la_i.
    \end{split}
\end{equation}
Then we obtain with the use of (\ref{cnoidal4}), (\ref{2-6}), and
(\ref{2-7}) the following expressions:
\begin{equation}\label{4-1}
\begin{split}
    \langle\mu\rangle&=\frac1L\oint\mu d\theta=\frac1L\oint\mu\frac{d\theta}{d\mu}d\mu=
    \frac1L\oint\frac{\mu d\mu}{2\sqrt{-P(\mu)}}=-\frac2L\frac{\prt I}{\prt s_2},\\
    \langle\mu^2\rangle&=\frac1L\oint\mu^2 d\theta=\frac2L\frac{\prt I}{\prt s_1}\\
    \langle\mu^3\rangle&=\frac1L\oint\mu^3 d\theta=-\frac{I}L+s_1\langle\mu^2\rangle
    -s_2\langle\mu\rangle+s_3,
    \end{split}
\end{equation}
where $I$ is a known integral
\begin{equation}\label{4-2}
\begin{split}
    I&=\int_{\la_2}^{\la_3}\sqrt{(\la_3-\mu)(\mu-\la_2)(\mu-\la_1)}\,d\mu\\
    &=\frac4{15}(\la_3-\la_1)^{5/2}[(1-m+m^2)E(m)-(1-m)(1-m/2)K(m)],
    \end{split}
\end{equation}
$K(m)$ and $E(m)$ being the complete elliptic integrals of the first
and second kind, respectively. The derivatives of $I$ with respect
to $\la_i$ are also known table integrals (Gradshtein $\&$ Ryzhik
1980):
\begin{equation}\label{4-3}
    \begin{split}
    \frac{\prt I}{\prt\la_1}&=-\frac12\int_{\la_2}^{\la_3}
    \sqrt{\frac{(\la_3-\mu)(\mu-\la_2)}{\mu-\la_1}}d\mu\\&=
    -\frac13\sqrt{\la_3-\la_1}[(\la_2+\la_3-2\la_1)E-2(\la_2-\la_1)K],\\
    \frac{\prt I}{\prt\la_2}&=-\frac12\int_{\la_2}^{\la_3}
    \sqrt{\frac{(\la_3-\mu)(\mu-\la_1)}{\mu-\la_2}}d\mu\\&=
    -\frac13\sqrt{\la_3-\la_1}[(\la_3-\la_1)K+(\la_1+\la_3-2\la_2)E],\\
    \frac{\prt I}{\prt\la_3}&=\frac12\int_{\la_2}^{\la_3}
    \sqrt{\frac{(\mu\la_2)(\mu-\la_1)}{\la_3-\mu}}d\mu\\&=
    \frac13\sqrt{\la_3-\la_1}[(2\la_3-\la_1-\la_2)E-(\la_2-\la_1)K].
    \end{split}
\end{equation}
We can easily express the $s_i$-derivatives in terms of $\la_i$
derivatives by differentiation of the formulae (see (\ref{2-7}))
\begin{equation}\label{4-4a}
    s_1=\la_1+\la_2+\la_3,\quad s_2=\la_1\la_2+\la_1\la_3+\la_2\la_3,\quad
    s_3=\la_1\la_2\la_3
\end{equation}
and solving the linear system for differentials. Simple calculation
gives
\begin{equation}\label{4-5a}
    \frac{\prt\la_i}{\prt s_k}=\frac{(-1)^{3-k}}
    {\prod_{j\neq i}(\la_i-\la_j)}.
\end{equation}
Then, combining (\ref{4-3}) and (\ref{4-5a}), we obtain the
derivatives $\prt I/\prt s_i$ and hence the expressions
\begin{equation}\label{4-6a}
\begin{split}
    &\frac{I}L=\frac2{15}(\la_3-\la_1)\left[(s_1^2-3s_2)
    \frac{E}K-\frac12(\la_2-\la_1)(\la_2+\la_3-2\la_1)\right],\\
    &\frac1L\frac{\prt I}{\prt s_1}=\frac16\left[2s_1\frac{E}K+
    s_1\la_1+\la_1^2-\la_2\la_3\right],\\
    &\frac1L\frac{\prt I}{\prt s_2}=-\frac12\left[(\la_3-\la_1)
    \frac{E}K+\la_1\right].
    \end{split}
\end{equation}
To complete the calculation of the right-hand side of
(\ref{pertmod}), we need also expressions
\begin{equation}\label{5-1a}
    \begin{split}
    \frac{L}{\prt L/\prt\la_1}&=2(\la_2-\la_1)\frac{K}E,\\
    \frac{L}{\prt
    L/\prt\la_2}&=-\frac{2(\la_3-\la_2)(1-m)K}{E-(1-m)K},\\
    \frac{L}{\prt L/\prt\la_3}&=\frac{2(\la_3-\la_2)K}{E-K}.
    \end{split}
\end{equation}
Collecting all contributions into perturbations terms, we obtain the
Whitham equations in the form
\begin{equation}\label{5-2a}
    \frac{\prt\la_i}{\prt T}+v_i\frac{\prt\la_i}{\prt X}=
    \rho_i=C_i[F(T)A_i-G(T)B_i],
\end{equation}
where $C_j$, $A_j$, $B_j$ and $v_j$, $j=1,2,3$ are specified by
formulae (\ref{5-3}) - (\ref{5-5}).

\end{document}